\documentclass[sigconf]{acmart}

\usepackage{float}
\usepackage{subfigure}
\AtBeginDocument{%
  \providecommand\BibTeX{{%
    \normalfont B\kern-0.5em{\scshape i\kern-0.25em b}\kern-0.8em\TeX}}}
\usepackage{tabularx}
\usepackage{bbding}
\usepackage{graphicx}
\usepackage{caption}
\usepackage{geometry}
\usepackage{amsmath}
\usepackage{makecell}
\usepackage{float}
\usepackage{color}
\usepackage{booktabs}
\usepackage{tabu}
\usepackage{hyperref}
\usepackage{xcolor}
\usepackage{appendix}

\copyrightyear{2025}
\acmYear{2025}
\setcopyright{acmlicensed}\acmConference[CHI '25]{CHI Conference on Human Factors in Computing Systems}{April 26-May 1, 2025}{Yokohama, Japan}
\acmBooktitle{CHI Conference on Human Factors in Computing Systems (CHI '25), April 26-May 1, 2025, Yokohama, Japan}
\acmDOI{10.1145/3706598.3713390}
\acmISBN{979-8-4007-1394-1/25/04}

\begin{document}

\title[Becoming My Own Audience]{"Becoming My Own Audience": How Dancers React to Avatars Unlike Themselves in Motion Capture-Supported Live Improvisational Performance}

\author{Fan Zhang}
\email{grazhang@cityu.edu.hk}
\orcid{0009-0009-9990-8820}
\affiliation{
\department{Studio for Narrative Spaces}
\institution{City University of Hong Kong}
\city{Hong Kong SAR}
\country{China}}

\author{Molin Li}
\email{lareinalml@gmail.com}
\orcid{0009-0002-7769-2300}
\affiliation{
\department{General Education and Research Unit}
\institution{The Hong Kong Academy for Performing Arts}
\city{Hong Kong SAR}
\country{China}}

\author{Xiaoyu Chang}
\email{changxiaoyu0527@gmail.com}
\orcid{0000-0002-1411-985X}
\affiliation{%
\department{Studio for Narrative Spaces}
  \institution{City University of Hong Kong}
  \city{Hong Kong SAR}
\country{China}}

\author{Kexue Fu}
\email{kexuefu2-c@my.cityu.edu.hk}
\orcid{0000-0002-2929-2663}
\affiliation{%
\department{Studio for Narrative Spaces}
  \institution{City University of Hong Kong}
  \city{Hong Kong SAR}
\country{China}}

\author{Richard William Allen}
\email{rwallen@cityu.edu.hk}
\orcid{0000-0003-2826-0990}
\affiliation{%
\department{School of Creative Media}
  \institution{City University of Hong Kong}
  \city{Hong Kong SAR}
\country{China}}

\author{RAY LC}
\authornote{Correspondences should be addressed to LC@raylc.org}
\email{LC@raylc.org}
\orcid{0000-0001-7310-8790}
\affiliation{
\department{Studio for Narrative Spaces}
\institution{City University of Hong Kong}
\city{Hong Kong SAR}
\country{China}}

\renewcommand{\shortauthors}{Zhang et al.}

\begin{abstract}
The use of motion capture in live dance performances has created an emerging discipline enabling dancers to play different avatars on the digital stage. Unlike classical workflows, avatars enable performers to act as different characters in customized narratives, but research has yet to address how movement, improvisation, and perception change when dancers act as avatars. We created five avatars representing differing genders, shapes, and body limitations, and invited 15 dancers to improvise with each in practice and performance settings. Results show that dancers used avatars to distance themselves from their own habitual movements, exploring new ways of moving through differing physical constraints. Dancers explored using gender-stereotyped movements like powerful or feminine actions, experimenting with gender identity. However, focusing on avatars can coincide with a lack of continuity in improvisation. This work shows how emerging practices with performance technology enable dancers to improvise with new constraints, stepping outside the classical stage.
\end{abstract}


\begin{teaserfigure}
\vspace{-0.3cm}
    \centering
    \includegraphics[width=1\linewidth]{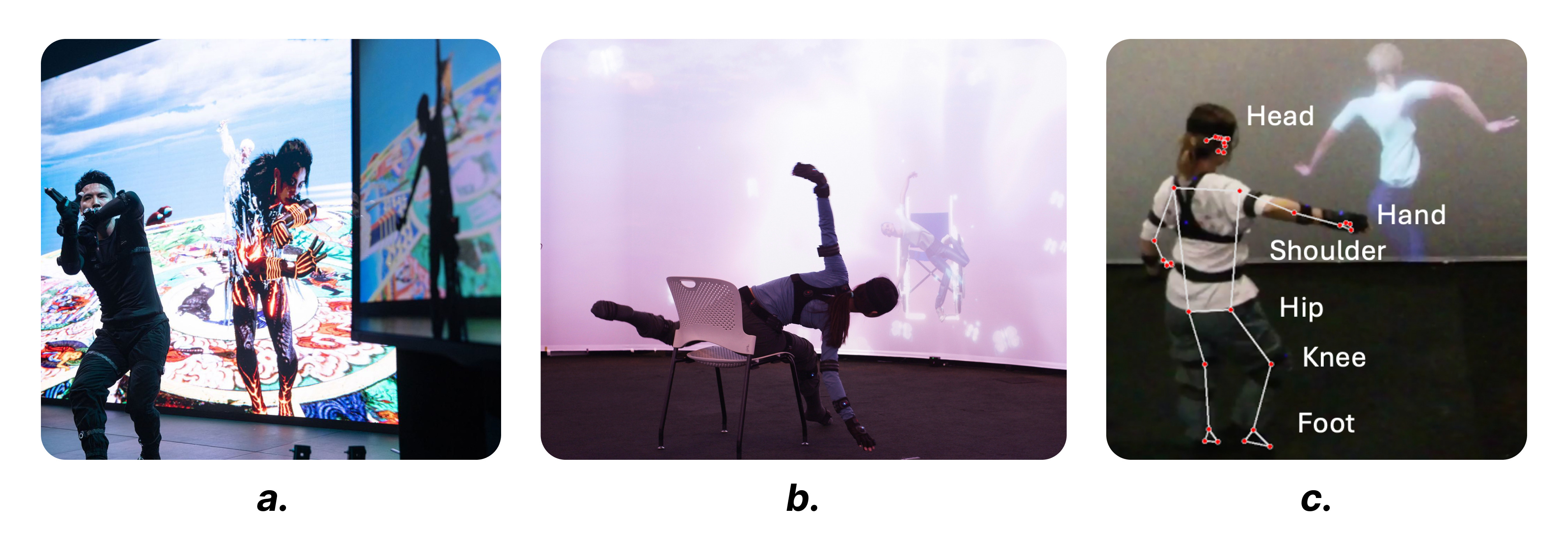} 
    \vspace{-0.7cm}
    \caption{Motion capture technology supported live performance and improvisation: \textbf{a.} Live performances with complex scenery that cannot physically take place \cite{noauthor_luyang_nodate}; \textbf{b. }Dancers did the improvisational performance with the avatar in a wheelchair (P6); \textbf{c.} Dancers' movement data were analyzed using computer vision (P9).}
    \label{fig:intro}
\end{teaserfigure}

\maketitle

\section{Introduction}\label{sec:Introduction}
\begin{figure*}[htbp]
 \vspace{-0.3cm}
  \centering
    \includegraphics[width=.99\linewidth]{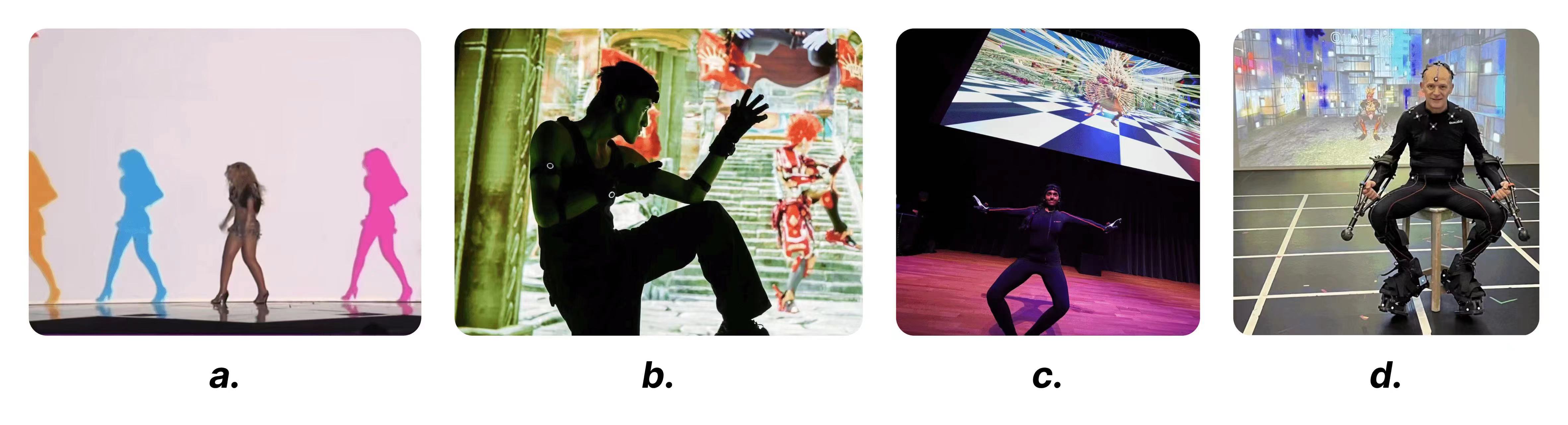}
  \vspace{-0.5cm}
  \caption{Performances with MoCap: \textbf{a.} Beyoncé used MoCap in 'Run the World (Girls)' at the 2011 Billboard Music Awards \cite{beyonce_beyonce_2011}; \textbf{b-d.} MoCap-supported live performance  \cite{noauthor_luyang_nodate,noauthor_maanasa_nodate,noauthor_studios44mocaplab_nodate}.}
 \label{artist}
  \vspace{-0.3cm}
\end{figure*}

In recent years, motion capture (MoCap) technology has been incorporated into live performances, where performers' motion-captured data is used to control an avatar on a digital version of the stage. MoCap's integration into real-time applications such as live streaming \cite{lu_i_2019,lu_streamsketch_2021}, VTubing \cite{wan2024investigating}, and phygital experiences \cite{meador_mixing_2004,andreadis_real-time_2010,demers_participative_2024}—where physical and digital experiences blend—allows performers to enact a scenario in a form of virtual storytelling. Unlike traditional dance performances, which are rooted in the physicality of the human body \cite{strutt2021virtual}, MoCap enables dancers to extend their movements into digital spaces, creating dance pieces performed by avatars on virtual stages \cite{noauthor_luyang_nodate,noauthor_zelia_nodate,noauthor_mocapdancing_nodate} (see Fig. \ref{artist}). For example, a recent multimedia art piece combines virtual reality, 3D animation, and MoCap to create an alternate digital universe with the artist's avatar \cite{noauthor_luyang_nodate}. This allows the artist to customize a virtual experience that would be impossible to recreate physically, enabling the telling of complex stories in dynamic environments.

In these new workflows supported by MoCap, how do dancers improvise their movements? How do dancers perceive this interaction and adjust their workflows? Improvisation is a critical part of dancers' daily practices, allowing them to push creative boundaries \cite{carlson2019shifting}. In digital spaces, dancers improvise with avatars of different sizes or genders, reflecting on how the visualizations will enable them to actually see the thing they could previously only imagine in their mental imagery \cite{raheb_choreomorphy_2018}. Through embodied interactions with avatars, performers gradually formed a relationship with avatars, experimenting with the way of moving in an unfamiliar body \cite{raheb_choreomorphy_2018}. MoCap-assisted co-design systems have shown the possibility of supporting artistic creation and movement exploration beyond conventional ways of moving during improvisation \cite{zhou_here_2023}.

However, research has shown that dancers often slip into habitual movement patterns, which can limit their improvisational creativity \cite{hagendoorn2003dancing}. To address this challenge, some researchers and choreographers have suggested making familiar things appear strange or new to prompt fresh perspectives within specific scenarios \cite{carlson2016ah}. This strategy, known as defamiliarization, disrupts habitual ways of thinking or moving, enabling dancers to create new choreographic materials by distancing themselves from their ingrained habits and past experiences \cite{carlson2019shifting}.

As dancers venture into virtual worlds, digital tools like avatars are being applied for improvisation and have been found to be particularly impactful as defamiliarization tools \cite{carlson2011scuddle}. Avatars offer a unique opportunity for dancers to step into new characters and detach from their usual physical identities, helping them view their bodies and movements from fresh perspectives \cite{zhou_here_2023,zhou_movement_2022}. Beyond dance, avatars have been widely used in social virtual reality and other virtual worlds, where they allow various users (even people with disability) to engage with others in digital spaces and experiment with self-representation and identity \cite{zhang2022s,mack2023towards, zimmermann2023self}. In the dance world, however, although the use of avatars has been explored in dance creation \cite{raheb_choreomorphy_2018,strutt2021virtual,akbas2022virtual} and dance learning \cite{tsampounaris_exploring_2016,fitton2023dancing}, how different features of these digital embodiments may influence the dancer’s movement, improvisation, and overall perception of their performance remain understudied.

To address this gap, we proposed the following questions:

\textbf{RQ1:} \textit{How do the features of avatars affect the way dancers move in motion-capture-supported live performances?}

\textbf{RQ2:} \textit{How do dancers perceive their bodily interaction in an immersive space while performing as different avatars?}

To explore these questions, we created five distinct avatars and invited 15 non-disabled professional dancers to interact with these avatars in both practice and performance settings. Considering the evolving trend in the dance world, which calls for engaging a broader range of bodies \cite{albright2010choreographing} rather than adhering to standardized body shapes (i.e., slim, long-limbed, and transcending everyday movement) \cite{au1988ballet}, we selected heavyset, gender opposition and disabilities affecting upper or lower limbs as four key conditions in our study. We only included non-disabled dancers to see how these professional dancers move and perceive when they improvised with avatars that were not “normative.” This approach aims to challenge physical limitations and embrace imperfection from a dance perspective, allowing for greater inclusivity and less constrained creative exploration. As a result, the avatars used in this study were: avatar with the same gender as dancers (normative), heavyset avatar (C1), avatar with the opposite gender (C2), avatar in a wheelchair (C3), and avatar without arms (C4). Semi-structured interviews, self-reported survey data, and computer vision data were collected during the study. By doing so, we sought to explore how dancers adapt their movement and improvisation when interacting with avatars with varying physical features.

Our findings reveal that avatars enabled dancers to distance themselves from their usual ways of moving, exploring new ways of moving through differing physical constraints of avatars or technical challenges. The dancers experimented with gender-stereotyped movements, such as powerful or feminine actions, which provided them with a unique space to engage with and experiment with gender identities. However, focusing on avatar embodiment also introduced challenges, particularly in distracting dancers and breaking continuity during improvisation. We then propose avatar design guidelines for future research to address the distractive effects.
This work contributes to the ongoing discourse on how digital tools can transform artistic practices, using avatars to show dancers what movements they are capable of when they take constrained and often neglected perspectives. By exploring how avatars that do not look like the performers themselves affect their dance routines, we highlight the way that people could gain insights into perspectives they may ordinarily struggle to understand by imagination alone from avatars of underrepresented body types, fostering inclusiveness and innovation in movement practices.

\section{Background}\label{sec:Background}
\subsection{Motion Capture (MoCap) Systems}

MoCap technology, widely used in entertainment for capturing human body motions in games and movies, is now being explored to preserve intangible heritage like dance \cite{hachimura_dance_2002} and musical performance \cite{morris_collaborative_2022}. Early optical MoCap systems, such as Kinect and camera techniques, track movements using reflective markers \cite{carlson_moment_2015}. These include mixed reality (MR) systems for dance training that guide trainees with CG characters’ motions \cite{hachimura_prototype_2004,hoang_onebody_2016}, and tools like “Super Mirror” \cite{marquardt_super_2012} and “YouMove” \cite{anderson_youmove_2013} that use mirrors and Kinect for real-time feedback. Recent advancements in MR mirrors have improved embodiment and spatial awareness for dance \cite{zhou_here_2023,zhou_movement_2022}. However, optical systems face challenges with lighting and portability, leading to the development of more portable inertial MoCap systems. These systems, like a wearable motion capture setup for outdoor use \cite{vlasic_practical_2007} and applications in dance learning \cite{tsampounaris_exploring_2016} and improvisation \cite{raheb_choreomorphy_2018}, expand the technology’s versatility and effectiveness in various environments.

While MoCap systems have come a long way since their inception, offering a more portable and user-friendly experience, how inertial MoCap systems can be utilized in dance improvisation and how they will influence dancers’ perception is still largely underexplored. Therefore, our RQ1 set out to explore how MoCap-generated avatars would affect the way dancers move in dance improvisation.

\subsection{Techniques Facilitating Dance Improvisation and Choreography}


Dance improvisation, often seen as the spontaneous creation of unplanned movements \cite{blom1988moment}, requires dancers to engage with their environment, emotions, and cultural contexts in real-time \cite{albright2003taken, potter2018beauty,trajkova2024exploring,torrents2015creativity}. Pioneers like Isadora Duncan advocated for natural, “free movements” that harmonize body, mind, and nature, transmitted primarily through oral tradition \cite{andrei_panov_duncan_2016,duncan_danse_2003,nameche_savoir-danser_2019,bang_designing_2023}. In contrast, William Forsythe’s \textit{Improvisation Technologies} provides a more structured approach with video tutorials, guiding dancers in exploring the body-space relationship through tasks like “creating points and lines in space” \cite{forsythe_william_2012}. Together, these perspectives highlight the diverse methods and philosophies that have shaped the practice of dance improvisation.


Research has shown that dancers often fall into habitual movement patterns, limiting improvisation creativity \cite{hagendoorn2003dancing}. To counter this, defamiliarization has been proposed as a tool for breaking these patterns \cite{bang_designing_2023, loke2013moving, wilde2017embodied, carlson2016ah}. Techniques like the “quick shift” help diversify movement by altering modalities \cite{kirsh2011creative}, while some researchers developed a framework combining human and system agency to enhance collaboration through HCI-inspired defamiliarization \cite{carlson2011scuddle,carlson2019shifting}. Merce Cunningham also employed defamiliarization by introducing randomness into his choreography \cite{alaoui2019making}, allowing dancers to explore unexpected creative possibilities \cite{carlson2016ah}. These approaches highlight defamiliarization as a powerful strategy for expanding the creative potential of dance improvisation.

Adding creative constraints is another effective way to counter habitual movement repetition, encouraging dancers to explore new patterns and possibilities. By setting boundaries, choreographers can drive more diverse and innovative improvisation. Balthasar Beaujoyeulx, for instance, used geometric patterns and numerical principles in his choreography, such as in \textit{Le Balet comique de la Royne (1581)} \cite{warburton2018imagine}. These constraints created a structured, organized choreography that reflected geometric ideals \cite{carter1992number}, demonstrating how imposed limitations can unlock fresh creative potential in dance.

Many technical tools have evolved from foundational frameworks to support dance creation. Rudolf Laban’s Laban Movement Analysis \cite{laban_modern_1963} has been instrumental in analyzing and teaching dance \cite{fdili_alaoui_seeing_2017,alaoui_choreography_2014}. Building on this, tools like the Choreographer’s Notebook \cite{singh_choreographers_2011,carroll_bodies_2012}, robotic movement systems \cite{dong_dances_2024, lc_presentation_2022, lc_active_2023}, and Knotation \cite{ciolfi_felice_knotation_2018} have been developed to document and organize choreographic ideas. These advancements highlight the ongoing evolution of tools that enhance and streamline the creative process in dance.

Although previous research has extensively explored techniques and technical tools to support dance improvisation and creation, there is still little knowledge on how the MoCap technique will influence dancers’ movement and improvisation based on previous frameworks. Therefore, our RQ1 aimed to explore how the MoCap-generated avatars will influence dancers’ movement.

\subsection{Stereotypes and Underrepresented Bodies in Dance}

Differences in body structure often lead to perceptions that men and women move differently in dance. Men typically have greater upper-body strength~\cite{janssen2000skeletal}, while women generally exhibit more hip flexibility~\cite{hufschmidt2015physical}. These physical differences, combined with cultural tendencies, have led to distinct gendered dance styles: masculine styles emphasize power and strength, while feminine styles highlight the flexibility and emotional expression~\cite{oliver2017introduction}. However, contemporary choreographers increasingly challenge these stereotypes by creating gender-neutral choreography. Techniques like role reversal and regendering, as exemplified by Pina Bausch in works like Carnations and Bandoneon, blur traditional gender boundaries and question conventional notions of masculinity and femininity~\cite{bellusci_lima_gender_2014}.

Moreover, there is an evolving trend in dance from a narrow ideal of standardized body shapes (the “sylph,” slim, long-limbed, and transcending everyday movement \cite{au1988ballet}) to embracing more natural and expressive forms of movement \cite{mcgrath2012dancing,foster1986expressionist}. Despite these changes, Western theatre dance remained physically exclusive until the 1960s, when pedestrian movement and contact improvisation emerged, creating opportunities for a broader range of bodies to participate and challenge the traditional, elite standards of dance \cite{albright2010choreographing}. 

Considering gender stereotypes and standardized body shapes in dance, we selectively chose heavyset, gender opposition, and physical constraints (wheelchair and no arms) as four conditions opposing traditional norms, including underrepresented groups in digital spaces.

\subsection{Avatar Mediated Interaction}

Avatars are integral to digital environments like live streaming \cite{freeman2024my,wan2024investigating}, virtual performances \cite{meador_mixing_2004,andreadis_real-time_2010,demers_participative_2024}, and social virtual reality (VR) \cite{fu2023mirror}, bridging physical and virtual communication \cite{friedl2002online,freeman2020my}. In terms of self-presentation, they contribute to body ownership \cite{guterstam2013invisible} and embodiment \cite{gonzalez2018avatar}, reflecting users’ movements and strengthening the connection between physical bodies and digital personas \cite{ducheneaut2009body,schroeder2001social}. Goffman’s performance metaphor suggests self-identity is shaped through interactive processes \cite{goffman1978presentation}, which avatars facilitate by simulating face-to-face interactions like self-monitoring \cite{lc_machine_2020} and mirrors \cite{fu2023mirror}.

Recent advances in MoCap technology and digital screens are challenging the traditional use of mirrors for visual feedback. MoCap systems and digital screens offer deeper insights into embodied experiences by refining dancers’ body schemas and performances through real-time feedback \cite{fitton2023dancing,radell_my_2014,preston_owning_2015}. These technologies bridge dancers and their avatars, enhancing the understanding of embodiment in dance \cite{hsueh_understanding_2019,aymerich-franch_second_2016}.

Furthermore, research shows how MoCap systems allow avatars to become a natural extension of inner self \cite{freeman2020my,gonzalez2018avatar}, influencing users’ perception, emotions and behavior \cite{yee2006walk,yee2009proteus,jones2017disrupting,kilteni2013drumming}. Research shows that greater similarity and realism between avatars and users improve the sense of embodiment \cite{latoschik2017effect}, while differing avatars can alter emotional responses and behaviors \cite{radiah2023influence}. For example, people act more intimately and more confidently when using more attractive avatars~\cite{fu2023mirror,messinger2008relationship}, across various embodiment types from static images to virtual reality \cite{chang2024investigating,van2013proteus,banakou2016virtual}. Avatars can also help mitigate stereotypes \cite{freedman2021effect} by allowing users to experience roles outside their usual identities, such as embodying different races \cite{banakou2016virtual}, genders \cite{radiah2023influence}, or disabilities \cite{mack2023towards,zhang2022s}. 

Although much research has focused on avatars in dance, little is known about how changes in avatar form—such as gender or body structure—affect self-perception and creativity in dance. Therefore, we aim to fill this gap by answering RQ1 and RQ2.

\section{Methods}\label{sec:Methods}
\subsection{Participants}

We conducted the experiment with 15 non-disabled professional dancers (12 female, 3 male, average BMI: 19.44, 7 out of 15 were classified as underweight). 7 participants were recruited from three local academic dance universities and 2 participants were recruited from a local dance club. We did a snowball sampling from these participants to see if they could recommend other professional dancers, and we recruited 6 additional participants. These dancers had different specializations, genres, and varied levels of experience in dance, as detailed in Table \ref{demographic}. 

Each dancer was equipped with a complete set of MoCap equipment and individually participated in a dance workshop within a 360-degree immersive environment. The study was approved by the institutional review board for human subject testing. Prior to participation, each dancer provided informed consent, acknowledging that their survey, interview data, and videos during the workshop would be collected anonymously for analysis and that they could withdraw from the study at any time for any reason. 

\raggedbottom
\begin{table*}[h!]
\centering
\caption {\label{tab:table1} Demographic Information of Participants}
\resizebox{\linewidth}{!}{
\begin{tabular} {ccccccc}
\toprule
\textbf{Participant}  &\textbf{Gender} &\textbf{Age}  &\textbf{Citizenship}   &\textbf{Dancing Experience}  &\textbf{Dance Genre}&\textbf{Specialization}   \\
\midrule
$P1$   & Male     & 23    &Canada & Expert   & Ballet &Choreography \\
$P2$   & Female   & 22    &Mainland China & Expert     &Standard Dance& Dance Performance\\
$P3$   & Female   & 25    &Mainland China & Moderate    &Contemporary Dance & Dance Performance\\
$P4$   & Female   & 22    &Hong Kong & Expert     &Modern Dance & Dance Performance\\
$P5$   & Female   & 21    &Hong Kong & Expert      &Modern Dance & Choreography\\
$P6$   & Female   & 26   &Mainland China & Expert   &Modern Dance & Choreography\\ 
$P7$   &  Female  & 20  &Mainland China  & Expert   &Modern Dance & Choreography\\
$P8$   &  Female  & 21  &Mainland China  & Expert   &Chinese Dance & Dance Theory\\
$P9$   & Female   & 23   &Mainland China & Moderate  &Hip Hop & Dance Performance\\
$P10$   & Female  & 21   &Hong Kong & Expert         &Modern Dance & Choreography\\
$P11$   & Female  & 20   &Mainland China & Expert &Modern Dance & Dance Performance\\
$P12$   & Female  &  32  &Mainland China & Expert & Chinese Dance & Dance Pedagogy \\
$P13$   & Female  &  47  &Hong Kong & Expert  & Swing Dance & Dance Performance \\
$P14$   & Male  &  25 &Mainland China & Moderate & Modern Dance & Dance Theory\\
$P15$   & Male  &   34 &Taipei & Moderate  & Blues Dance & Dance Performance \\
\bottomrule
\multicolumn{7}{c}{Moderate: dancing experience 1.5 to 5 years, Expert: dancing experience > 5 years.}\\
\end{tabular}
}
\label{demographic}
\Description{Demographic Information of Participants}

\end{table*}

\subsection{Workshop Process}

\subsubsection{Procedure}

\begin{figure*}[htbp]
 \vspace{-0.1cm}
  \centering
    \includegraphics[width=.95\linewidth]{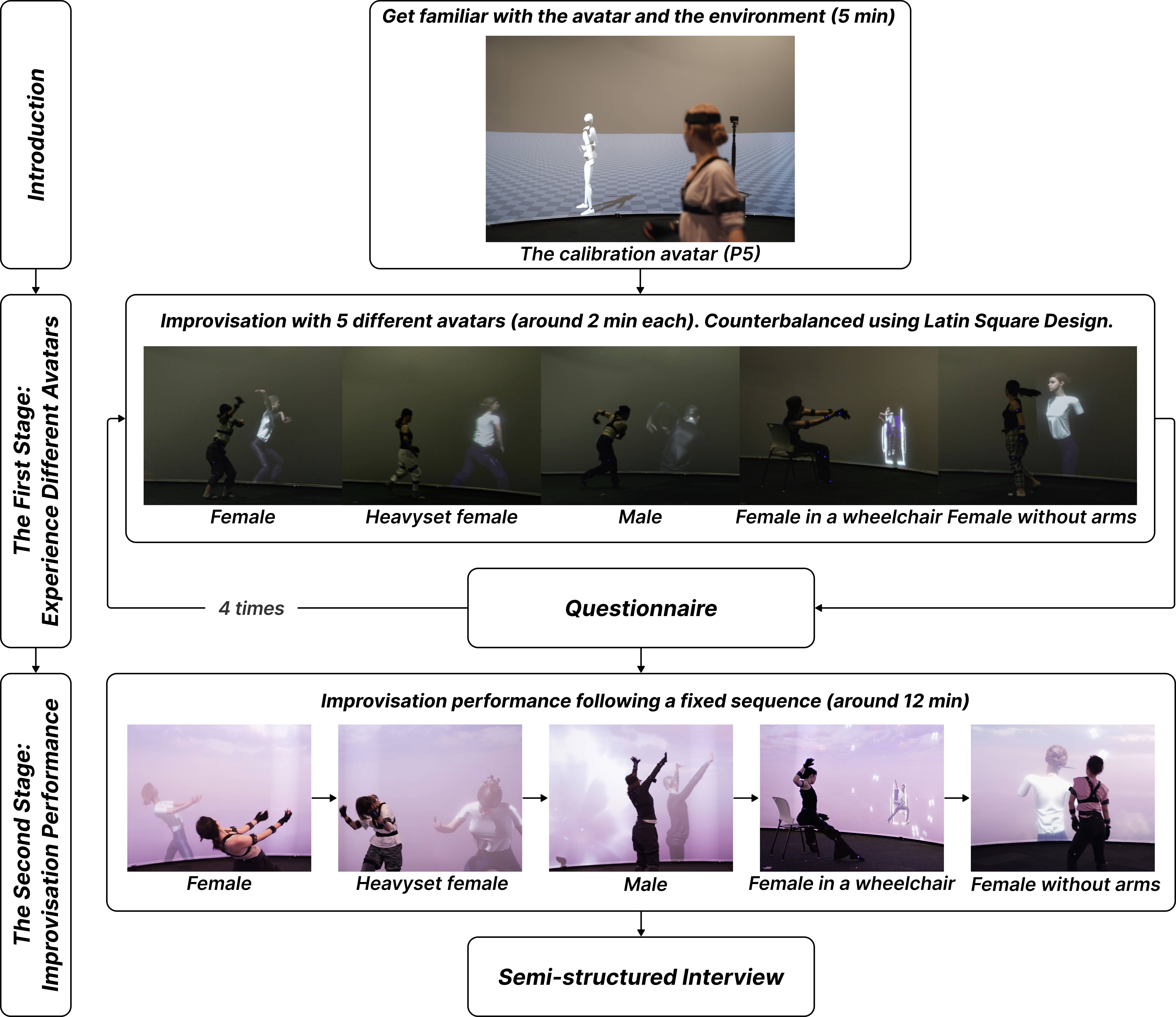}
  \vspace{-0.3cm}
  \caption{The workshop process uses avatars that female participants experienced as examples. Male participants experienced avatars of the opposite gender (male, heavyset male, female, male in a wheelchair, and male without arms).}
 \label{study_flow}
  \vspace{-0.3cm}
\end{figure*}

Each workshop lasted approximately three hours (see Fig. \ref{study_flow}). At the beginning of each session, a dancer wore a MoCap suit and spent 5 minutes familiarizing themselves with the default avatar. During the first stage of the workshop, the dancer improvised with five different avatars against a black background to minimize the influence of the virtual environment and other external factors. In the second stage, the dancer performed an improvisation sequence using the five avatars in a predetermined order. During both stages, we used the same music for improvisation. The music we selected was instrumental, with no lyrics, and had a relatively steady rhythm. It was intended to serve as background music that would not overly influence dancers’ movements or choices. Dancers were not asked to mimic the avatars or pretend to be avatars, but they were allowed to do whatever they wanted based on what they saw and heard.

During the first stage of the workshop, female dancers experienced five different avatars (female, heavyset female, male, female in a wheelchair, and female without arms). They improvised with each for 1.5 - 2.5 minutes based on their preferences. Male dancers experienced avatars of the opposite gender. The sequence of avatars in the first stage was balanced using Latin Square design. After the improvisation with each avatar, participants were asked to fill in the adapted embodiment questionnaires  \cite{gonzalez2018avatar} and think aloud about their improvisation experience with the avatar. To minimize external influences, the avatars were presented against a black background, excluding the virtual environment and other factors from affecting the dancers’ responses.

During the second stage, each participant was asked to do an improvisation performance (around 12 minutes) with a fluid experience of all the avatars (each lasted 1.5 - 2.5 minutes, according to the dancer’s preference). The sequence of avatars that female participants experienced was female, heavyset female, male, female in a wheelchair, and female without arms. Male dancers experienced avatars of the opposite gender. Participants were told the performance concept before the performance, as detailed in Section \ref{story}.

After experiencing all the avatars and the performance, the dancer was invited to a semi-structured interview regarding the improvisation experience with different avatars and perception of body and space in the 360-degree immersive environment. 

\subsubsection{Performance concept} \label{story}

The performance narrative was designed to simulate a dancer’s dream, where they encounter a virtual life characterized by surreal transformations and challenges, navigated through a series of metaphorical portals. One researcher told the dancer the narrative before the performance as follows: 

The dancer awakens in an unfamiliar environment and begins to explore through improvisation. A portal appears, leading them to the next dream, where they find they have gained 50 pounds but continue to dance. A second portal causes their gender to change, and they improvise with this new identity. Stepping through a third portal, they discover their legs are paralyzed and perform from a wheelchair while interacting with butterflies. The fourth portal restores their leg function but leaves them without arms, prompting new interactions with the butterflies. The dream concludes with the appearance of the final portal.

\subsection{Technological Implementations}

\subsubsection{Platform and performance area}

\begin{figure*}[htbp]
 \vspace{-0.3cm}
  \centering
    \includegraphics[width=.99\linewidth]{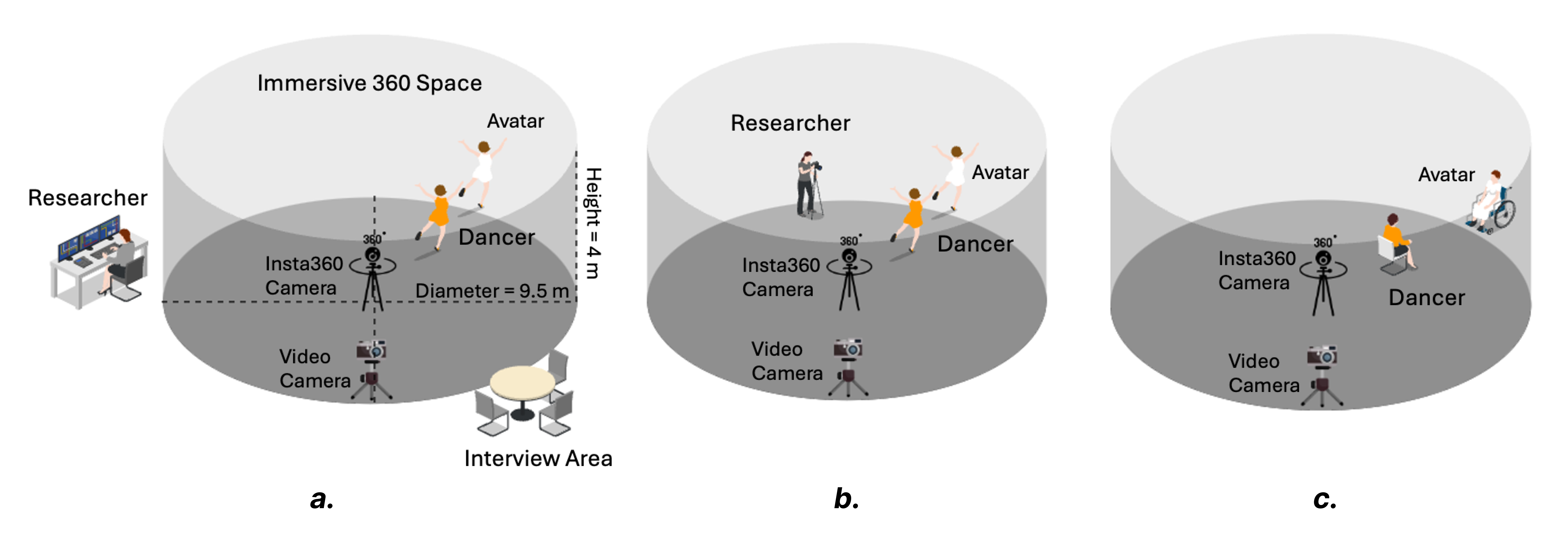}
  \vspace{-0.3cm}
  \caption{Floor plan and experimental setup: \textbf{a.} Only one dancer improvised with avatars in the environment during the first stage; \textbf{b.} One researcher took narrative videos besides the 360 camera and video camera during the second performance stage; \textbf{c.} When experiencing the avatar in a wheelchair, a normal chair was placed in the environment as a set.}
 \label{floorplan}
  \vspace{-0.3cm}
\end{figure*}

\begin{figure*}[htbp]
 \vspace{-0.3cm}
  \centering
    \includegraphics[width=.95\linewidth]{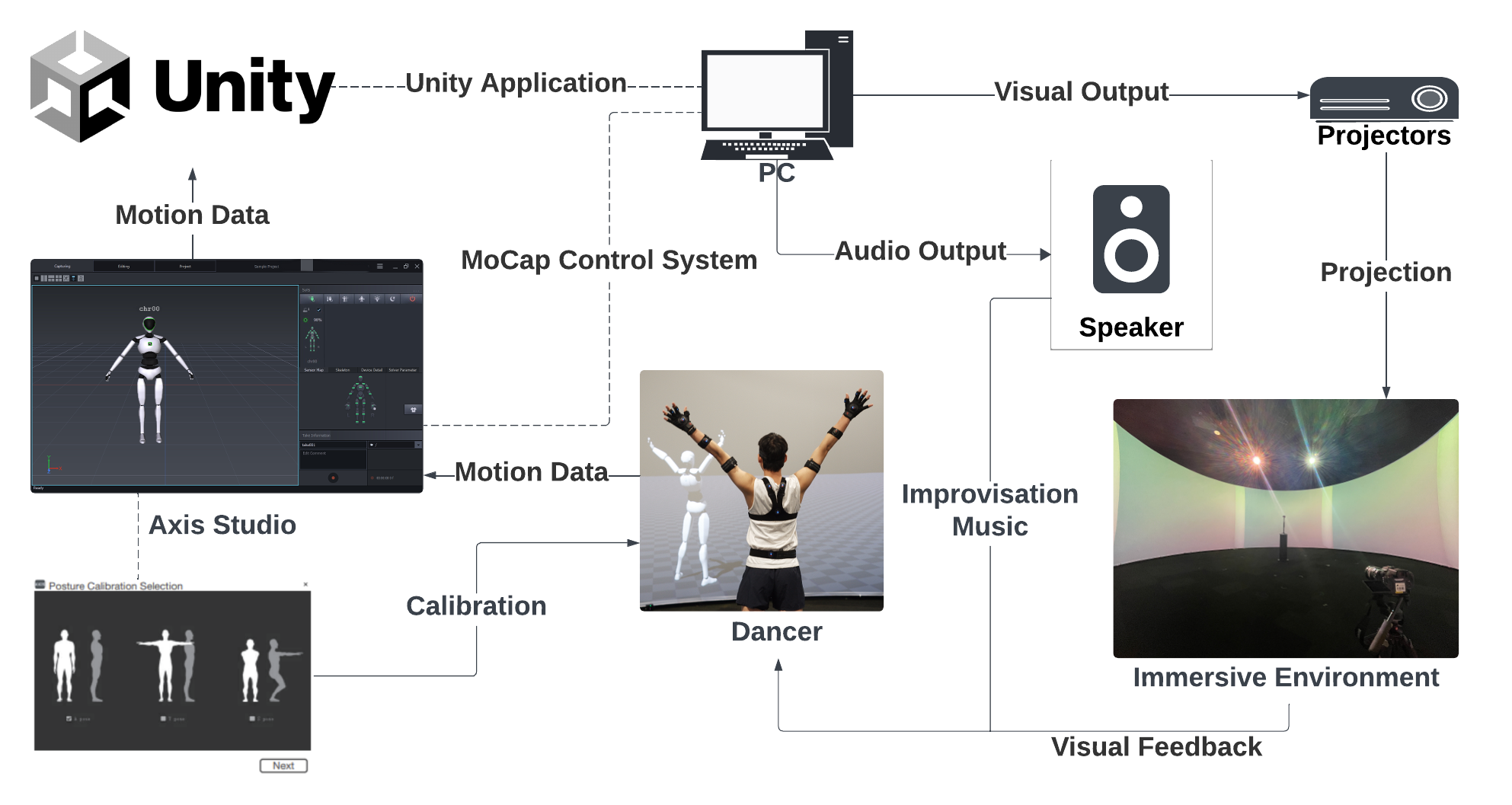}
  \vspace{-0.3cm}
  \caption{An overview of the system: Dancers wearing the MoCap suit did calibration and sent motion data through Axis Studio software connected to the PC. The motion data from Axis Studio was live-streamed to Unity applications with avatars and the virtual environment. The PC was connected to projectors and speakers for visual and audio output. Dancers could see real-time visual feedback and hear improvisation music simultaneously.}
 \label{system}
  \vspace{-0.3cm}
\end{figure*}

To facilitate real-time interaction between dancers and avatars from all angles of the performance venue, we conducted the experiment using a cylindrical projection screen (measuring 4 meters in height and 9.5 meters in diameter) in an immersive 360 space (see Fig. \ref{floorplan}). Dancers were equipped with a MoCap suit (Perception Neuron), which allowed their movements to be synchronized in real-time with the avatars in the virtual environment (see Fig. \ref{system}). These synchronized movements were projected onto the cylindrical screen. The virtual environment was developed in Unity, enabling us to design the scene in accordance with the narrative setting. 

During the workshop, a 360-degree camera was placed in the center of the space, and a video camera with a 20mm super wide angle was placed at the edge of the screen (see Fig. \ref{floorplan}). During the first stage, only one dancer improvised with different avatars in the immersive space. During the second stage, one researcher was aside taking narrative videos for the dancer. When dancers experienced the avatar in a wheelchair, a normal chair was placed in the environment as a set.

\subsubsection{MoCap calibration procedure}

To ensure accurate data collection, a calibration process was performed prior to each improvisation session of the first stage. Before each session, we checked that the sensors were in the right place. We performed calibration using Axis Studio, which involved a series of standardized poses to align the sensors with the dancer’s body (see Fig. \ref{system}). In the second stage, to ensure the integrity of the improvisation performance, we only performed calibration before it started and when recalibration was required, e.g., straps or sensors slippage due to intense movement or severe avatar distortion due to loss of synchronization with software.

\subsubsection{Mapping of Unity virtual environments to immersive 360 space}

The Unity virtual environment was designed to map the spatial positioning of avatars to an immersive 360 space, ensuring a coherent and intuitive user experience. 

\begin{figure*}[htbp]
 \vspace{-0cm}
  \centering
    \includegraphics[width=.99\linewidth]{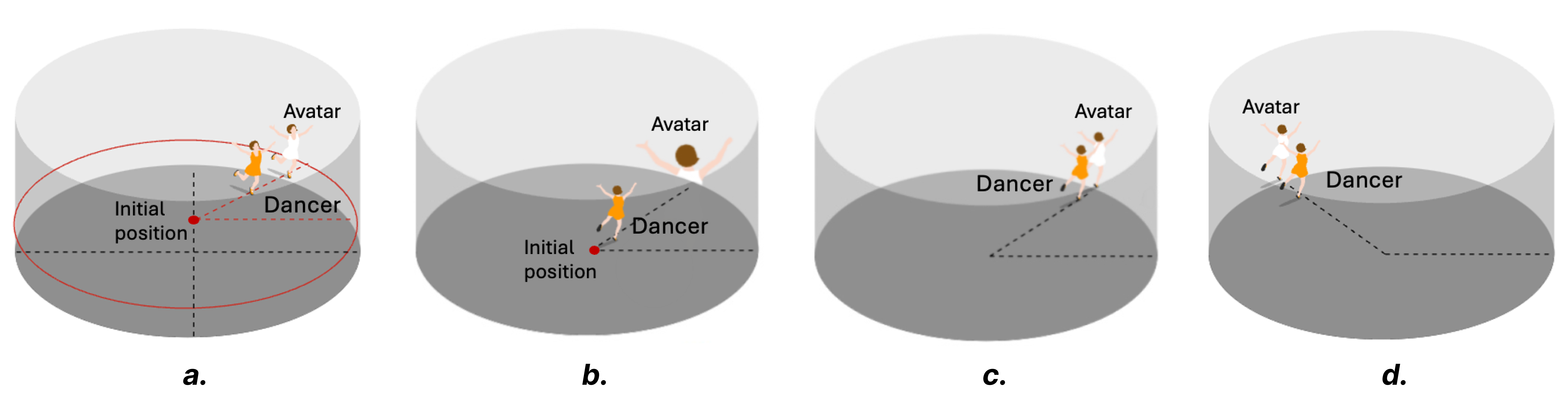}
  \vspace{-0.3cm}
  \caption{Mapping of Avatars to the environment: \textbf{a.} The initial position of the avatar (experienced by P1-P6) was set to a forefront position to capture videos better; \textbf{b.} The initial position of the avatar (experienced by P7-P15) was set to the center of the space, where the avatar appeared at its largest and could only be partly seen; \textbf{c.} When dancers stood close to the screen, the avatar was the smallest as life-size; \textbf{d.} The avatar was made to face the same direction as the dancer, except for the avatar in a wheelchair.}
 \label{mapping}
  \vspace{-0.3cm}
\end{figure*}

To better capture videos of dancers dancing with avatars, we first set the initial position of avatars to a forefront position in the physical space (see Fig. \ref{mapping} a). The first six dancers experienced this mapping relationship and called for a more apparent 360-degree interaction with avatars. Therefore, the remaining nine dancers experienced avatars with the initial position at the center of the virtual environment, corresponding to the most extensive view of avatars on the screen (see Fig. \ref{mapping} b). When dancers stood close to the screen, the avatar appeared at its smallest, matching the life-size proportions (see Fig. \ref{mapping} c). The avatars generally aligned with the dancer’s orientation, with both facing the same direction (see Fig. \ref{mapping} d), just as the way many performers and dancers have experienced with MoCap \cite{noauthor_luyang_nodate,noauthor_zelia_nodate,noauthor_mocapdancing_nodate}. An exception was the experience with the avatar in a wheelchair, which was made to be face-to-face for better interaction. This positional and relational mapping was consistent throughout the immersive 360 environment, ensuring that the avatar’s representation accurately reflects the dancer’s movements and positioning, regardless of their location.

\subsubsection{Avatars and virtual environment}

\begin{figure*}[htbp]
 \vspace{0.3cm}
  \centering
    \includegraphics[width=.95\linewidth]{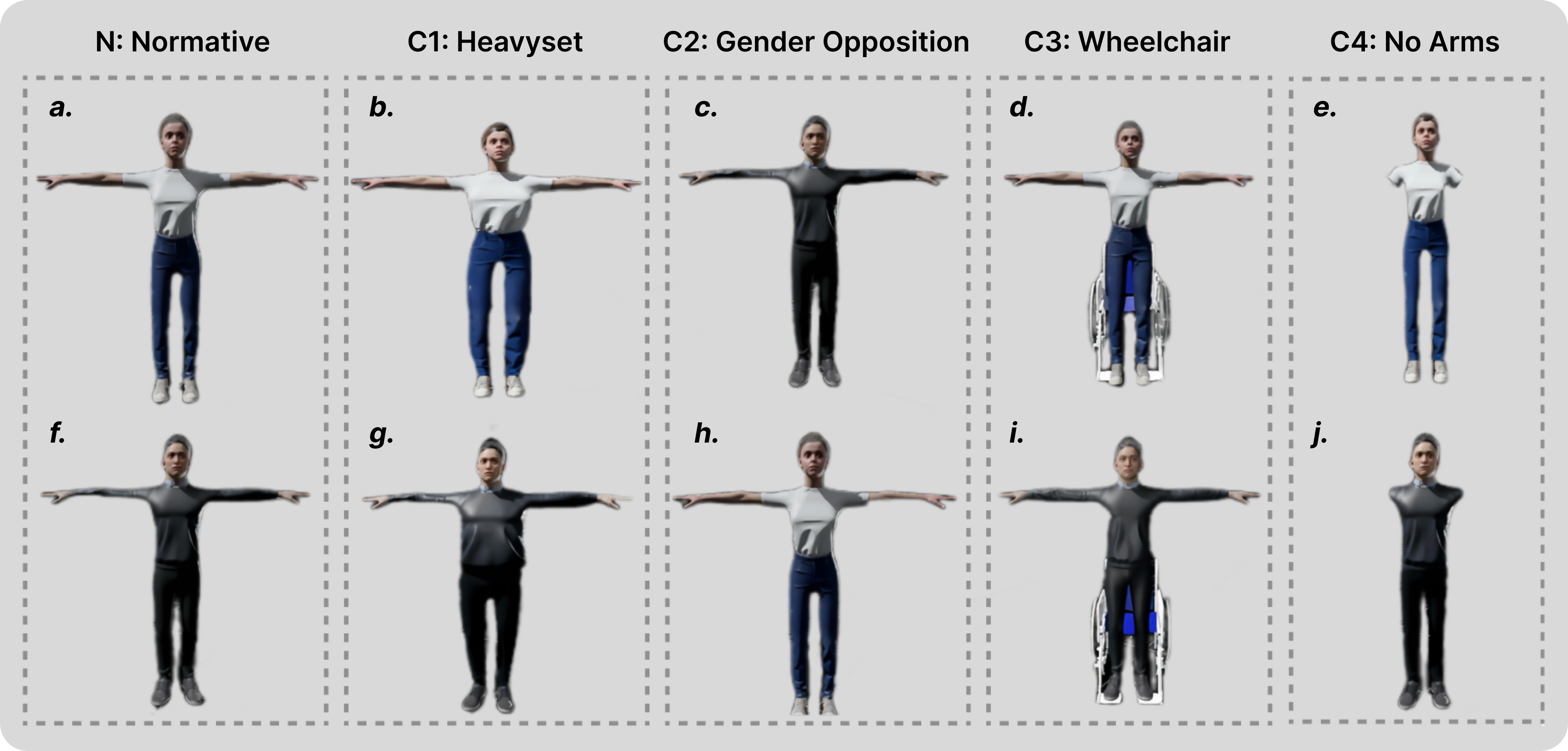}
  \vspace{-0.3cm}
  \caption{Avatars used in the workshop included normative condition and 4 conditions (heavyset, gender opposition, wheelchair, and no arms): The normative condition was the avatar with the same gender as dancers. \textbf{a - e.} were avatars that female dancers experienced,  \textbf{f - j.} were avatars that male dancers experienced.}
 \label{avatar}
  \vspace{0cm}
\end{figure*}

\begin{figure*}[htbp]
 \vspace{0.3cm}
  \centering
    \includegraphics[width=.95\linewidth]{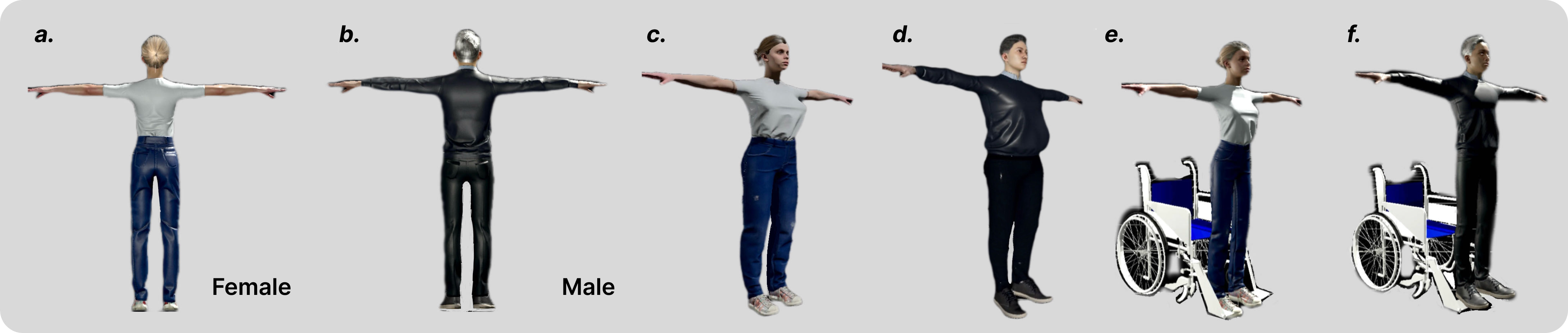}
  \vspace{-0.3cm}
  \caption{The female avatar could be identified by the updo from the back, white top, and blue jeans, while the male avatar had short hair and black clothes:  \textbf{a - b.} Back view of N (normative);  \textbf{c - d.} Side view of C1 (heavyset);  \textbf{e - f.} Side view of C3 (wheelchair).}
 \label{side view}
  \vspace{0cm}
\end{figure*}

Avatars used in the workshop were customized using Unity assets \cite{noauthor_character_nodate}. The avatars were designed to test the impact of different affordances (i.e., weight, gender, and physical disability) on dancers’ improvisation while controlling for other variables (see Fig. \ref{avatar}). To ensure that the study focused on the selected variables, all avatars were kept at the same height, dressed in similar clothing, and exhibited consistent ethnicity and overall appearance. The female avatar could be identified by the updo from the back, white top, and blue jeans, while the male avatar had short hair and black clothes (see Fig. \ref{side view}). The normative avatar had body shapes similar to the participants, which could be a bit underweight. The heavyset avatar was stretched in width to show the idea of gaining weight. The avatar in a wheelchair was made by adding a wheelchair to the normative avatar. The avatar without arms was made by disabling the upper limb skeleton in Unity.

\begin{figure}[htbp]
 \vspace{-0.3cm}
  \centering
    \includegraphics[width=.99\linewidth]{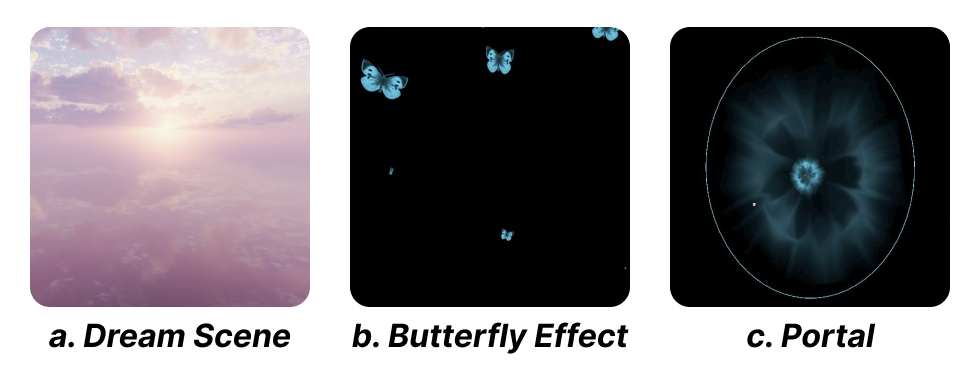}
  \vspace{-0.4cm}
  \caption{Virtual environment: \textbf{a.} The environment used in the performance; \textbf{b.} Butterfly effect used for the performance of the avatar in a wheelchair and the avatar without arms; \textbf{c.} The portal used in the performance.}
 \label{envir}
  \vspace{-0.3cm}
\end{figure}

During the first stage, the environment was made all-black without introducing confounding variables. During the second stage, the environment was made with gentle lightness and dreamy colors, with butterflies and portals, to fit the performance concept (see Fig. \ref{envir}).

\subsection{Data Collection and Analysis}

\subsubsection{Semi-structured interview and Think-aloud method}

After each improvisation session, dancers were encouraged to express their feelings and thoughts when dancing with different avatars (using the Think aloud method \cite{charters_use_2003}). Participants were asked after interacting with each avatar whether they noticed the change of avatar and what the change was as an awareness check. After the improvisation with five avatars and the improvisation performance, the dancers each participated in semi-structured interviews conducted by one researcher and one note-taker. The interviews focused on dancers’ improvisation experience with different avatars generated by motion capture in the immersive 360 space and their perceptions of five different avatars (see detailed interview questions in Section \ref{interview}).

The transcripts of the interviews were initially coded by two researchers. Then, two researchers revised the codes and discussed the themes with the whole research team. The process was repeated iteratively until the team reached a final consensus \cite{braun_using_2006,lungu_coding_2022}.

\subsubsection{Computer vision analysis}

We placed an Insta360 camera and a video camera with a 20mm wide angle in the immersive space to record videos of each session. The videos used for computer vision were those from the first stage of the workshop, eliminating the influence of confounding variables such as virtual environments. 

The computer vision analysis for the dancers was computed using 2D pose estimation from video frames. Specifically, key body landmarks such as the hips, shoulders, and knees were extracted from each frame using the MediaPipe Pose framework. The average of these joint coordinates approximated the dancer’s Center of Gravity (COG) and joint movements, representing a balanced point of the body during movement. The COG data was tracked across time, comparing its variation between different conditions, such as N and C1. The average joint movement was tracked across time, comparing their variations between N and C3, N and C4. This analysis allowed us to visualize the vertical displacement of the COG and average joint movement across the dance performance, providing insight into how the avatar’s physical characteristics influenced the dancer’s movement.

\subsubsection{Survey}

After experiencing each avatar, dancers were presented with the adapted version of the standardized embodiment questionnaire (7-point Likert-scale) \cite{gonzalez2018avatar} to quantitatively evaluate their subjective experiences about body ownership, agency, control, and self-location (tactile sensations and response to external stimuli were excluded based on relevance) during their improvisation (see Section \ref{survey question}). The scores were calculated as suggested in the questionnaire \cite{gonzalez2018avatar}. We ran \textit{Paired Samples T Test} to measure different conditions (C1 - C4) compared with normative condition (N). We also compared each question and did \textit{Paired Samples T Test}.

\section{Results}\label{sec:Results}

\subsection{Influence of Avatars on Movements (RQ1)}

In this section, we present how the features of avatars affect the way dancers move during their improvisation. Dancers were found to distance themselves from habitual movements (Section \ref{defamiliarization}), experiment with gender-stereotyped movements, navigate new physical constraints, and create new movements responding to technical constraints of MoCap itself (Section \ref{creativityfromconstraints}). However, their movements could sometimes be disrupted if they focused too much on avatars (Section \ref{distraction}). Dancers also mentioned they could potentially invest less effort in movement qualities due to the limited expressiveness of avatars (Section \ref{limited expressiveness}).

\subsubsection{Facilitating natural improvisation with different avatars through defamiliarization} \label{defamiliarization}


Dancers had a consensus of natural improvisation, although many of them had difficulty improvising naturally as professional dancers. Our participants reported that avatars recovered their sense of improvisation as "defamiliarization" tools that provide interesting stimulation and new improvisation materials, leading dancers to distance themselves from their own habitual movements or repeated movements due to the music. 

Dancers reported using improvisation as part of their daily routine and experience, though they sometimes felt restricted as professional dancers. For P2, improvisation was \textit{"more of a way of life instead of a technical thing for dancers"}.  She believed that improvisation was the natural expression of emotions in life and also helped to communicate with others non-verbally. She believed that improvisation could, in some way, provide more information and bring people closer. P6 confirmed this opinion by saying, \textit{"People are born knowing how to dance."} P5 emphasized that improvised movements should be based on the scene and feelings at the moment. However, sometimes professional dancers had difficulty doing improvisation naturally, instead they easily fell into the atmosphere and rules of work. As P2 mentioned, improvisation was quite easy for her when she began dancing, but later, it became something she had to treat carefully when becoming a professional dancer. The virtual space and avatars, however, gave her a chance to get out of the rules.

\begin{quote}
    \small\itshape
"This may be the current situation of dancers...when you improvise, you will always fall into the atmosphere and rules of work. I think the biggest feeling that virtual space brings to me is that it inspires the impulse of dialogue and communication between me and the entire space or avatars...it makes me forget some of the original rules of dancers' improvisation." (P2)
    \end{quote}

\begin{figure}[htbp]
 \vspace{-0.3cm}
  \centering
    \includegraphics[width=.95\linewidth]{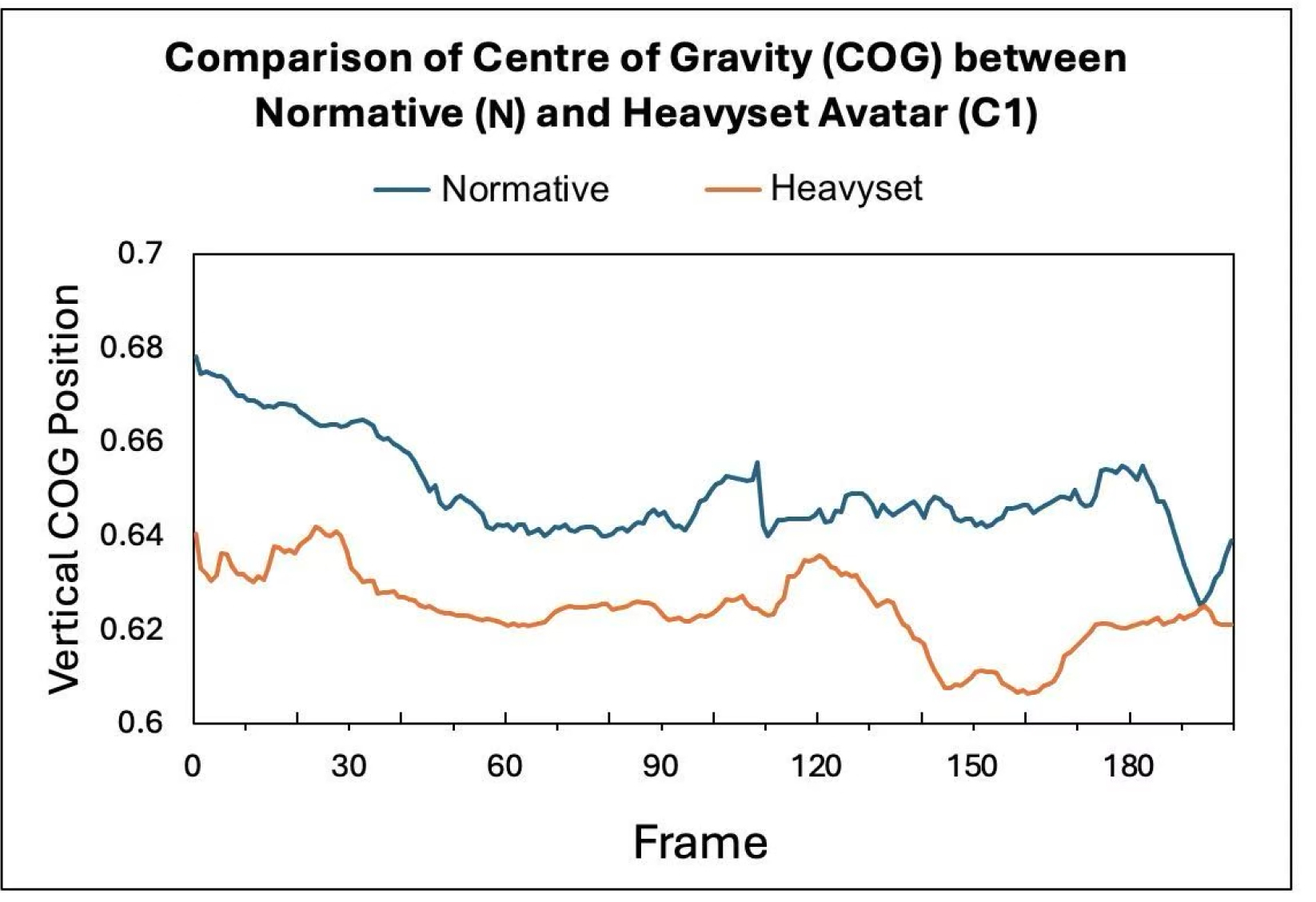}
  \vspace{-0.3cm}
  \caption{Defamiliarization effect of C1 (heavyset) suggested by computer vision results: Compared to N (normative), P9 lowered her center of gravity when dancing with C1.}
 \label{cv_cog}
  \vspace{-0.3cm}
\end{figure}

\begin{figure}[htbp]
 \vspace{0.3cm}
  \centering
    \includegraphics[width=.95\linewidth]{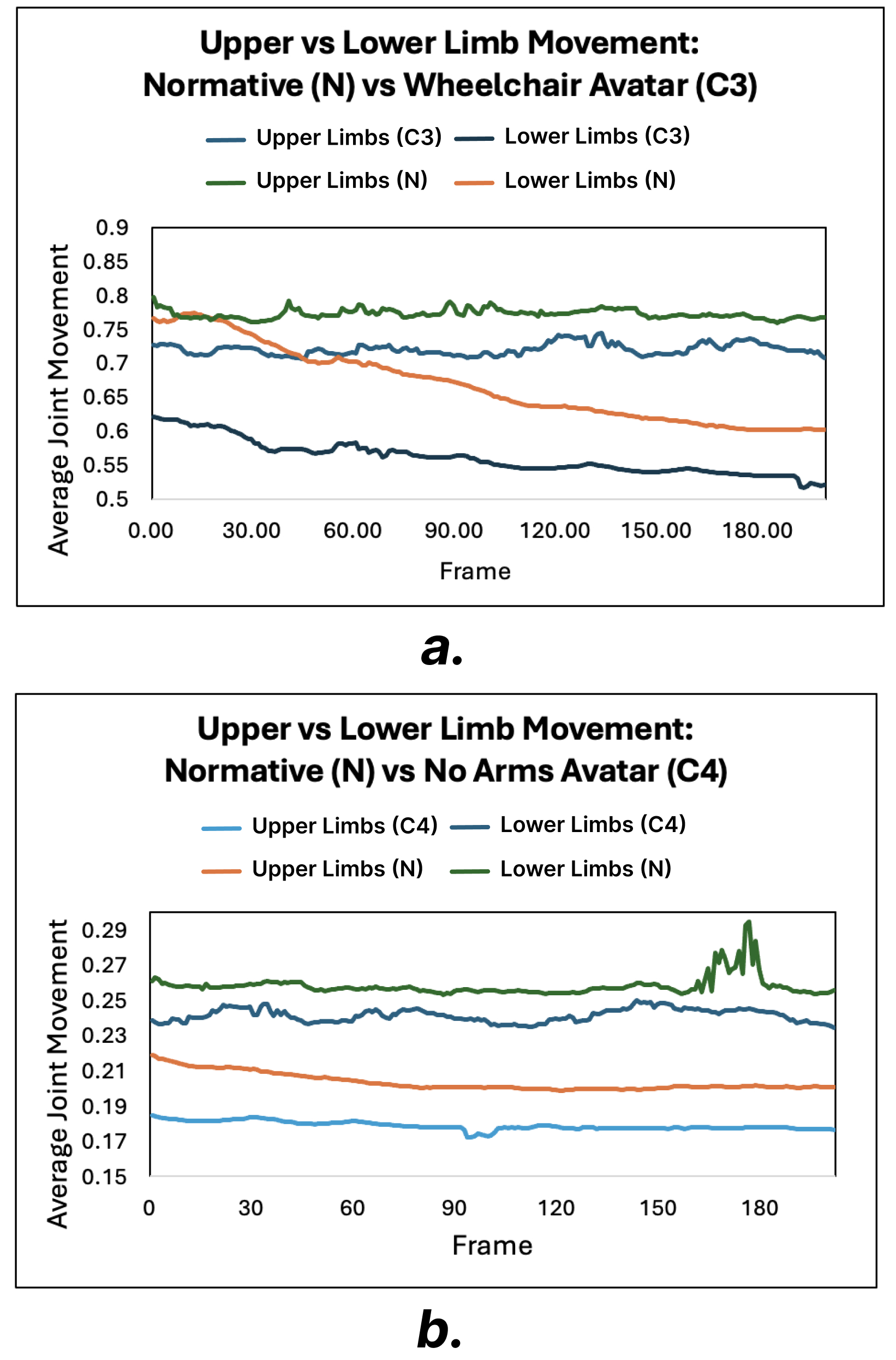}
  \vspace{-0.3cm}
  \caption{Defamiliarization effect of C3 (wheelchair) and C4 (no arms) suggested by computer vision results: \textbf{a.} Compared to N (normative), P15 restricted both his upper and lower limbs when dancing with C3; \textbf{b.} Compared with N, P13 restricted both her upper and lower limbs when dancing with C4.}
 \label{cv_limb}
  \vspace{-0.3cm}
\end{figure}

Dancers indicated that avatars provoked unfamiliar stimulation and  defamiliarized themselves, with varying effects on improvisation according to different affordances of avatars. Dancers reported that avatars' appearances made them think differently and influenced their ways of using their own bodies, thus changing their movements. Avatars that diverged more significantly from the dancers' own identities had a greater impact on altering their movements. For example, the normative avatar was considered similar to dancers, making them dance almost the same way as usual. The heavyset avatar made P1 feel his own body getting heavier and thus made heavier movements (see Fig. \ref{cv_cog}). Dancing with the opposite gender made dancers think and do their movements in a different way (see Section \ref{gender identity}). Disabled avatars gave dancers greater constraints on upper or lower limbs (see Fig. \ref{cv_limb}), but in some way gave them more materials and psychological \textit{"liberation"} to improvise with (see Section \ref{creativityfromconstraints}). Overall, these varied responses to different avatars highlighted how defamiliarization could both challenge and enrich dancers’ improvisation, enabling them to explore new movements and expressions.

Moreover, some dancers reported avatars led them to distance themselves from repeated and generally soft movements caused by the music. P5 specifically mentioned repeated movements due to the same music during the whole, while disabled avatars (C3 and C4) made her move differently and jump out of the movement habits. P9 did mostly soft, smooth movements like waves because of the music, while the male avatar made her do more powerful and strong movements. Consistent with P9, P12 also did more powerful male postures in Chinese classical dance when dancing with the male avatar, breaking the influence of soft music. This might suggest that avatars could serve as a catalyst for breaking habitual movement patterns, encouraging dancers to explore a wider range of physical expression.

\subsubsection{Creativity from constraints} \label{creativityfromconstraints}

Dancers found several constraints brought by avatars, including disabled affordances of avatars, lack of delicate movements, and distorted motion induced by technical constraints. Nevertheless, our participants noted the creativity brought by certain limitations, just as constrained improvisation in dance.

All the dancers noticed that the constraints brought by disabled avatars (i.e., avatars without arms or in a wheelchair), although limiting their use of certain body parts, gave them even more creative possibilities. According to P4, these constraints resembled constrained improvisation in dance, making purposes clearer and distancing dancers from movement habits. Similarly, P5 viewed disabled avatars as an instance of constrained improvisation that could point out the direction and \textit{"produce more interesting points"}. When improvising with the avatar in a wheelchair, dancers tended to limit their use of lower limbs and focused more on upper limbs and torso area (see Section \ref{self restrict}); when interacting with the avatar without arms, dancers intentionally restricted their own upper limbs and explored more on other body parts (see Section \ref{freedom}). Beyond just the movements, P2 and P3 also reported that disabled avatars provoked deeper thoughts on the feelings and emotions they would like to convey. For dancers, disabled avatars uncovered creative possibilities by giving clearer movement directions and deepening emotional expressions through certain constraints.

Dancers reported that using disabled avatars led them to put their emphasis naturally on unrestricted body parts, which is better than using props for constrained improvisation in their daily practice. Some participants noted using props (e.g., tying hands, putting something behind the knees, using elastic bands to restrain the thighs and calves, etc.) to constrain their practice to allow for improvisation. However, P8 mentioned that she would still want to move restricted body parts when using external props, while disabled avatars changed her mindset, giving her internal forces to explore movements only in unrestricted body parts. Similarly, P1, P13, P14, and P15 reported thinking deeply and exploring what the movement possibilities would be if they did not have upper or lower limbs. This might suggest avatars could provide dancers with clearer instructions and restrictions than props in improvisation by giving them internal forces.

\begin{quote}
    \small\itshape
"I think its limitation can bring me more possibilities... I am now a healthy person, I can assume that I don't have it, or I tie my hands... when I go to perform, in fact, I still want to move. But if you really don't have this (hand), it may not be able to move." (P14)
    \end{quote}

\begin{figure*}[htbp]
 \vspace{-0.3cm}
  \centering
    \includegraphics[width=.99\linewidth]{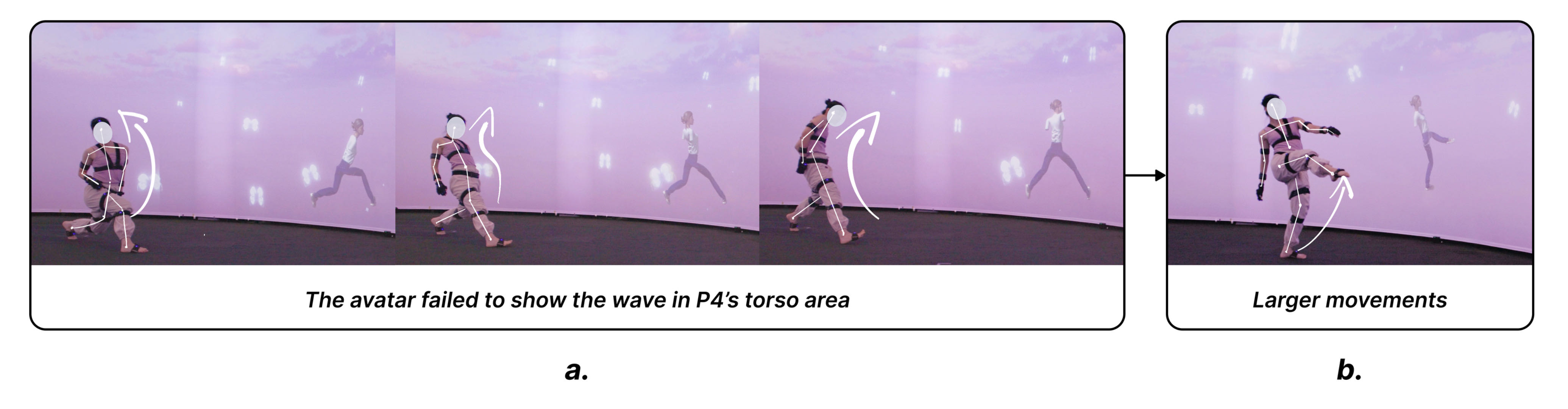}
  \vspace{-0.5cm}
  \caption{Dancers chose to do larger movements due to the lack of movement delicacy in avatars: \textbf{a.} P4 did a wave in her torso area, but the avatar failed to show it on the upper body; \textbf{b.} P4 chose to do larger movements, like lifting one leg instead.}
 \label{delicate}
  \vspace{-0.3cm}
\end{figure*}

\begin{figure*}[htbp]
 \vspace{0cm}
  \centering
    \includegraphics[width=.99\linewidth]{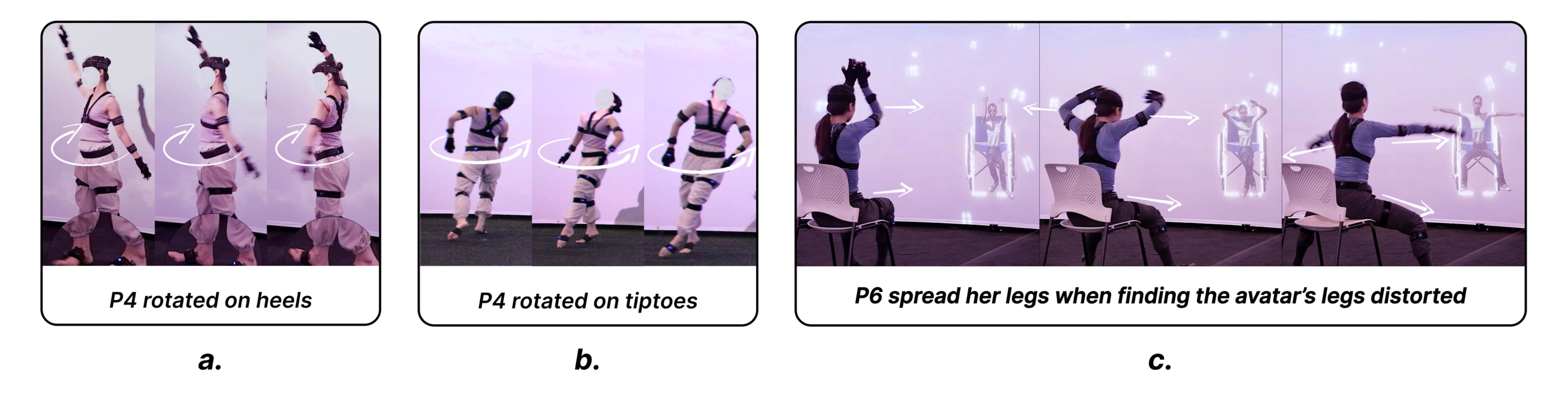}
  \vspace{-0.4cm}
  \caption{Adaptive movement sets:  \textbf{a - b.} P4 used heels and tiptoes to rotate due to straps on feet;  \textbf{c.} P6 found the avatar's legs distorted, thus spreading her legs, trying to make the avatar look reasonable.}
 \label{rotate}
  \vspace{-0.3cm}
\end{figure*}

Besides, avatars brought in other innovative opportunities or movement influences through technical constraints, such as failure to show delicate movements, the sense of restraint caused by the equipment, and movement distortion due to calibration error or position deviation. Many dancers (P3, P4, P5, P10, P11, P15) noticed that the movements of avatars were not delicate enough to show the trajectory of all the small joints, only capturing some relatively large movement paths, which made P4 and P5 choose to do larger movements (see Fig. \ref{delicate}). Dancers (P1, P4) reported adaptive movement changes due to the MoCap suit itself. To avoid moving the positions of sensors (especially those on feet), P4 switched to her heels to rotate and experienced weight shifting (see Fig. \ref{rotate} a - b). 
Some dancers (P6, P7, P9, P10) took the challenge of movement deformation in avatars as a chance for creativity and movement inspiration. 
Both P6 and P10 mentioned trying to straighten their legs to make the avatar look reasonable when seeing distorted legs in avatars (see Fig. \ref{rotate} c). P9 reported movement inspirations given by distorted avatars, as she would like to correct deformed body parts in avatars to normal. Overall, dancers experienced unfamiliar situations and drew novel inspirations even from the technological constraints of the MoCap system itself.

\begin{quote}
    \small\itshape
"If I look at myself in the mirror, I will still be looking at myself, and I will be in a fixed mode. But because of the avatar, I can find different angles. Because of its limitations, I can see my own limitations." (P6)
    \end{quote}

\subsubsection{Looking at one's own body instead of the outside environment} \label{distraction}

Many participants (except P3, P6, and P7) emphasized the importance of focusing on their own bodies instead of the outside environment in improvisation. External avatars or human partners, as well as themselves in mirrors, would distract dancers from perceiving their own body and make movements appear \textit{"flat."}

Dancers (P2, P5, P8, P11) found similar disadvantages of looking at avatars on the flat screen and themselves in mirrors. Mirrors, although frequently used in dance studios for learning movements or checking uniformity in group dance, could make movements appear "flat" in improvisation. As P2 said, when looking at movements through mirrors, dancers' eyes could be limited and unable to experience \textit{"multi-directional movements and different spaces."} P5 also indicated that movements would appear flat if dancers focus too much on one point (looking at a mirror or a screen), while more three-dimensional movements could be generated by looking inside. Dancers like P8 and P11 also reported choosing not to look at mirrors by closing their eyes to immerse themselves in dance. This suggests that similar to mirrors, viewing avatars on a flat screen can limit dancers' perception of movement, leading to a flattened experience. 

\begin{quote}
    \small\itshape
  "Because when you look in the mirror, you will not be able to dig inward to find out where the movement comes from, how it flows, its skeleton, how its path changes, where its force comes from, you may only care about its external shape. So many teachers may mention the need for looking inside, that is, while your eyes are observing everything around you, you need something like an endoscope to observe everything happening in your body." (P5)
    \end{quote}

It was frequently reported by dancers (P1, P4, P5, P9, P11, P13, P14) that they were distracted by the avatars due to the responsibility for avatars, resulting in failure to perceive their own bodies and movements. Dancers like P1 and P4 reported the feeling of responsibility for the avatar, being mindful about how the avatar would be seen on the screen. The focus on how the avatars appeared on screen led some dancers, like P13 and P14, to adjust or limit their movements for the sake of the avatar’s visual presentation. Others, like P5 and P9, struggled to connect with their internal sensations, with their attention diverted to the avatar’s external shape. For P11, this issue seriously impacted her movements, as she almost lost balance and fell down during the rotation due to the focus on the screen. This suggests that avatars might detract from dancers’ ability to remain connected with their own bodies, ultimately disrupting their movement and overall performance.

Furthermore, several participants (P2, P4, P12, P13, P14) noted a significant difference between improvising with avatars and human partners, often finding that their focus on the avatars caused them to lose concentration on their own bodies. P4 held the opinion that she would try not to overly consider her partner in modern dance training, as doing so could lead to a loss of self-awareness. She explained, \textit{"When we pay too much attention to the other person, we will lose ourselves, for sure."} This struggle was mirrored when dancing with avatars, where she felt a disconnection from her own body whenever she had to focus on the avatar, only to regain that sense of self when the avatar was no longer visible. Other dancers, although they felt that it was necessary to perceive and care about partners, acknowledged the importance of building connections with each other. Establishing connections with avatars, however, was even harder than with human partners due to the lack of physical sensations. This phenomenon was also found by P2, who indicated that the virtual figure distracted herself from her dance and her body, pulling her to the \textit{"outside space"} in a \textit{"playful"} way. Her comment of \textit{"putting 50\% of attention on the avatar and giving up 50\% of myself"} echoed the opinion of P4. This suggests that avatars could hinder dancers’ ability to maintain self-awareness and connection in their movements.

\begin{quote}
    \small\itshape
  "I also feel a little confused. Is the moving body following me? Or is this body the avatar's? It will affect the trajectory of my body to a certain extent. When I see its existence, I seem to lose something of myself." (P4)
    \end{quote}

\subsubsection{Limited expressiveness beyond movements in avatars} \label{limited expressiveness}

\begin{figure*}[htbp]
 \vspace{-0.3cm}
  \centering
    \includegraphics[width=.99\linewidth]{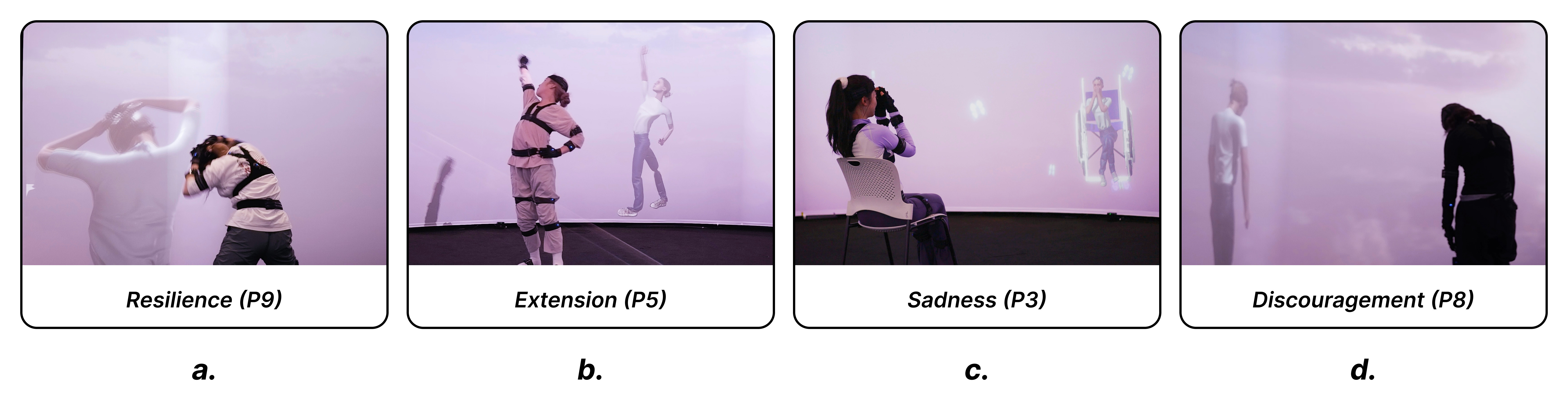}
  \vspace{-0.5cm}
  \caption{Avatars lacking in expressiveness: \textbf{a.} The resilience in P9's movement could hardly be seen in the avatar; \textbf{b.} P5's movement had the sense of extension while the avatar did not; \textbf{c.} P3 had strong sadness when facing the avatar in a wheelchair, which could be seen from her facial expression and body language, which the avatar did not show; d. The avatar was lacking in both the emotion and amplitude of movements when P8 tried to show discouragement.}
 \label{quality}
  \vspace{-0.3cm}
\end{figure*}

Dancers (P4, P5, P6, P8, P10, P12) noted that avatars exhibited limited expressiveness outside of just the movements themselves (see Fig. \ref{quality}). For instance, P5 could not change the way the avatar looked when she attempted to make a soft movement. It was still just the movement without expressiveness, failing to represent the tone and quality of the movement. This issue persisted when other dancers attempted to demonstrate other movement qualities such as the sense of extension, resilience, or large resistance. P8 reported that the avatar could not show different styles and qualities of various dance genres, especially diverse styles of Chinese ethnic folk dance. She used \textit{Anhui Huagudeng} as an example, saying that human dancers could make the audience \textit{"know at a glance which ethnic group it belongs to,"} while the avatar could not show any style. The avatar consistently displayed the \textit{"stiffness"} and in short of the \textit{"vividness"} that could be seen in human dancers, exhibiting limited expressiveness outside of just the movements. 

\begin{quote}
    \small\itshape
  "Ballet is very rigid, and modern dance tends to be softer, while the style of ethnic folk dance is diverse. For example, Korean Folk Dance focuses on the inner breath, not an outward one, so it is actually very difficult (for the avatar) to capture." (P8)
    \end{quote}

Moreover, dancers (P3, P8, P12, P13, P14) noted that avatars often failed to convey the emotions they intended to express. For instance, P3 displayed strong sadness through her facial expression and body language while dancing with the avatar in a wheelchair, but the avatar was unable to convey this emotion. Similarly, P8 performed with a sense of discouragement that was evident even from her back, yet the avatar could only replicate the physical movement, lacking emotional depth. This lack of emotional expressiveness in avatars was observed to potentially diminish the layers and richness of the dancers’ performances.

Nevertheless, some dancers like P6 saw potential benefits in this limitation, believing it could reduce the effort required to achieve a desired virtual image. She noted that she had opportunities to \textit{"be lazy,"} meaning she could obtain the virtual image she wanted with less emphasis on movement qualities. To capture a virtual image that showcased movement trajectories, she found that she didn't need to invest as much effort in refining movement qualities, such as resilience and extension, or in conveying emotional expression.

\subsection{Dancers' Perception While Performing as Different Avatars (RQ2)}

In this section, we present how dancers perceive their bodily interactions while performing as different avatars. Dancers felt the avatar in a wheelchair less resembled themselves and had less control over it (Section \ref{sec:embodiment}). Generally, dancers' perceptions were changed by the appearance of avatars, which then influenced their movement choices. For example, the gender shift changed dancers' thoughts internally and holistically, provoking their reflections on gender identity and experiments with gender-stereotyped movements (Section \ref{gender identity}). Disabled avatars also made dancers feel restrained and self-restricted their own body (Section \ref{self restrict}). Nevertheless, the avatar without arms brought a sense of freedom through avatar movement constraints, provoking deeper thoughts on lower body creative possibilities (Section \ref{freedom}). Additionally, dancers perceived avatars in two distinct ways: either as extensions of themselves, enabling them to observe their own movements from a new perspective, or as challenging partners that were harder to communicate with than their human counterparts. In the latter case, dancers often found themselves being led by unfamiliar or disabled avatars.

\subsubsection{Embodiment} \label{sec:embodiment}

During dancers' improvisation with avatars, they exhibited varied perceptions towards different avatars. We found that dancers felt the avatar in a wheelchair less resembled themselves and had less control over it. Besides, dancers felt the virtual body of the avatar of the opposite gender less likely to be their own body. We also found that their movements were significantly influenced by the avatar without arms.

\begin{figure*}[htbp]
 
  \centering
    \includegraphics[width=.99\linewidth]{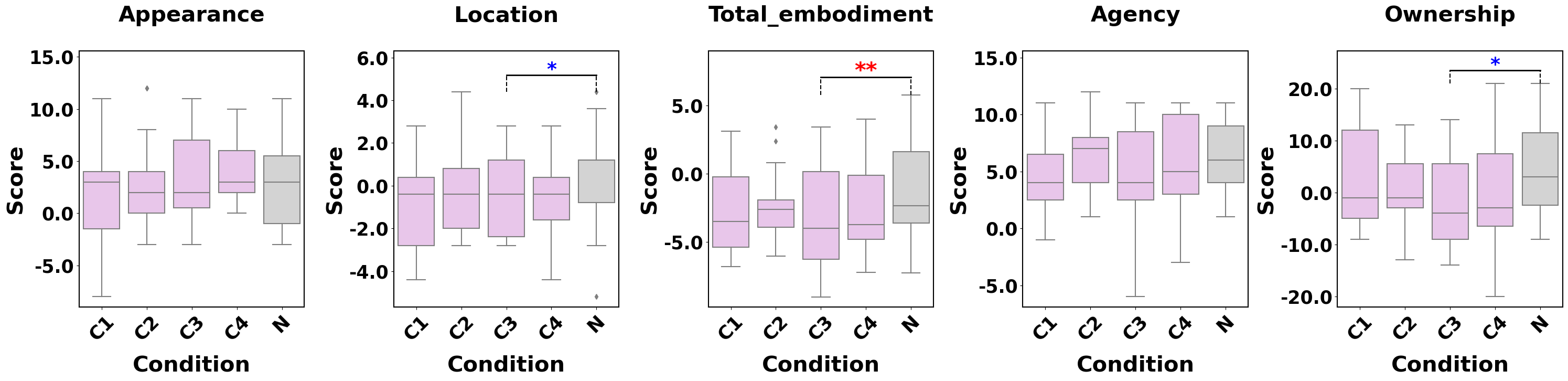}
  \vspace{-0.3cm}
  \caption{Measure of how well dancers felt avatars resembling themselves (** p < 0.05, * p < 0.1): There was a significant main value of C3 (wheelchair) on \textit{Total embodiment} (\textit{p} = 0.0221 < 0.05), a tending significance of C3 (wheelchair) on \textit{Location} (p = 0.0967 < 0.1) and \textit{Ownership} (p = 0.0606 < 0.1)}
 \label{survey}
  \vspace{-0.3cm}
\end{figure*}

We applied a \textit{Paired Samples T Test} to measure four conditions (C1 - C4) compared with the normative condition (N) (see Fig. \ref{survey}). There was a significant main value of C3 (wheelchair) condition on \textit{Total embodiment} (\textit{p} = 0.0221 < 0.05); the score difference of C3 (Mean = 3.9756) was significantly lower than that of the score difference in N (Mean = 7.4889). This means dancers felt the avatar in a wheelchair less likely to be themselves. This was consistent with P4's comment that the avatar in a wheelchair made her feel less authentic. There was a tending significance of C3 on \textit{Location} (p = 0.0967 < 0.1) and \textit{Ownership} (p = 0.0606 < 0.1), suggesting a potential trend that warrants further investigation.\\

We also compared each question and did a \textit{Paired Samples T Test} again (see Table \ref{question}). For example, there were significant effects of C2 (\textit{p} = 0.029 < 0.05) and C3 (\textit{p} = 0.021 < 0.05) conditions on \textit{Q1."I felt as if the virtual body was my body."}, meaning dancers felt the virtual body of avatar with opposite gender and avatar in a wheelchair less likely were their own body. The values of C3 on \textit{Q4."I felt as if the virtual body I saw when looking at the screen was my own body."} (\textit{p} = 0.008 < 0.05) and \textit{Q6."It felt like I could control the virtual body as if it was my own body."} (\textit{p} = 0.001) were statistically significant, meaning dancers felt the avatar in a wheelchair less resemble their own body and generally had less control over this avatar. The value of C4 condition on \textit{Q8."I felt as if the movements of the virtual body were influencing my own movements."} (\textit{p} = 0.037 < 0.05) was statistically significant, meaning dancers' movements were significantly influenced by the avatar without arms, which was consistent with their feedback in interviews.

\subsubsection{Exploring gender identity through avatar identification} \label{gender identity}

Most participants (except P5 and P15) felt the gender shift was a very different experience from other avatars, changing their internal thoughts and ways of moving. The improvisation experience gave them a chance to experiment with gendered movement, exploring their own gender identity in a less constrained environment.

Dancers (P1, P3, P7, P8, P9) reported a more internal and holistic change brought by the gender shift when compared to other avatars. Gender shift was considered as more of an internal change by P1, while other avatars gave external shape changes. Avatars with the same gender made P9 focus more on specific body parts being changed, while the gender shift gave her a totally different angle. She mentioned how she understood the avatar from a holistic perspective, \textit{"You can sense some of the personality traits from his appearance, some sense of power that he may have, and give him a story...instead of just observing what his head, hands, or feet look like."} Similarly, P3, P7, and P8 reported changed ways of thinking and \textit{"attitudes towards things"} while facing an avatar with the opposite gender, thus impacting their dance movements. This might suggest that gender shifts in avatars went beyond physical appearance, fostering deeper cognitive and emotional responses that influence improvisation and self-expression.

Interestingly, dancing with the opposite gender provoked P4 and P14's reflections on gender identity, giving them a chance to navigate and experiment with aspects of gender identity that might not be as freely explored in physical dance environments. P4 observed that while gender was often emphasized by others in the dance world, whether through symbolic clothing or the choice of male or female dancers for specific roles, her own perspective was different. She found herself questioning why gender held such significance in dance. For her, gender didn't define her artistic expression in dance, even though she acknowledged its importance in personal relationships and the formation of self. Similarly, P14, often perceived as a more feminine male, experimented with how different gender expressions influenced movement. The digital realm offered dancers a more fluid and less constrained environment for gender exploration and expression, allowing dancers to engage with and reimagine their gender identity.

\begin{figure*}[h!]
  \centering
    \includegraphics[width=.99\linewidth]{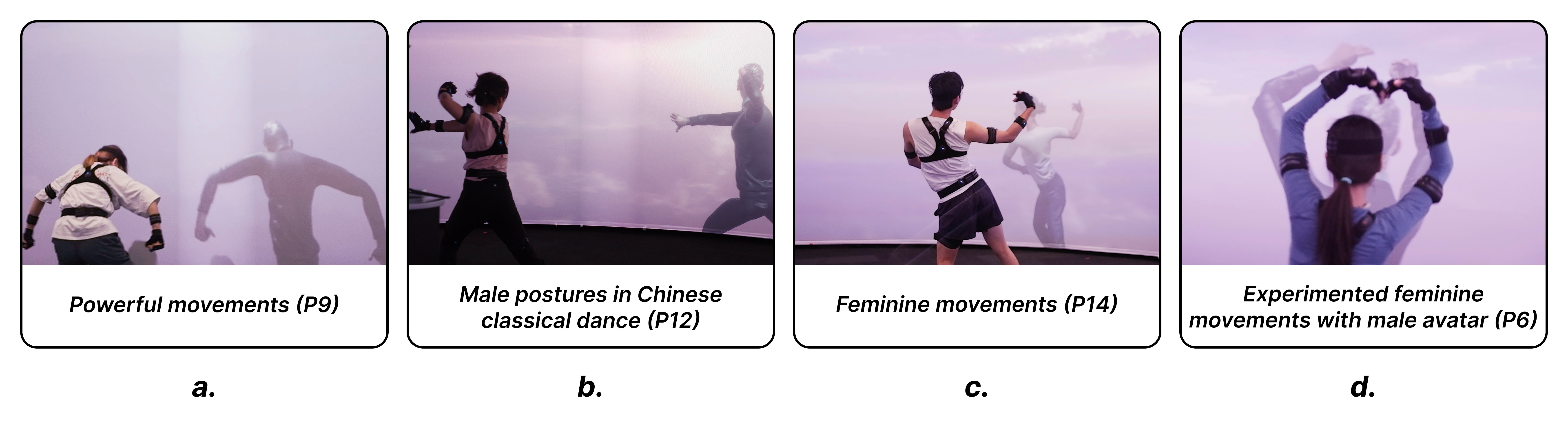}
  \vspace{-0.5cm}
  \caption{Gendered movement sets: \textbf{a.} P9 did powerful movements; \textbf{b.} P12 used a lot of male postures in Chinese classical dance; \textbf{c.} P14 did feminine movements through body curves; \textbf{d. }P6 deliberately experimented with some feminine movements with male avatars for fun.}
 \label{gender}
  \vspace{-0.3cm}
\end{figure*}

With the digital stage for gender exploration, dancers experimented with gendered movements. Dancers explored movements potentially with gender stereotypes, e.g., P4 and P9 mentioned \textit{"large, straight, and powerful movements"}; P8 did \textit{Yingge Dance} (a Chinese war dance), which was often done by male dancers to show the power; P12 chose to do male postures in Chinese classical dance; while P14 did more feminine movements with soft quality  (see Fig. \ref{gender} a-c). Meanwhile, some participants (P4, P6, P8) deliberately experimented with feminine movements with male avatars, observing the funny and humorous effect on avatars (see Fig. \ref{gender} d). Nevertheless, some dancers like P5 and P15 viewed dance as gender neutral, finding the gender shift made little difference in their improvisation. These varied responses highlight the diverse ways dancers explore gender identity through movement in virtual environments.

\subsubsection{Self-restricted movement sets due to disabled avatars} \label{self restrict}

\begin{figure*}[htbp]
 \vspace{0cm}
  \centering
    \includegraphics[width=.99\linewidth]{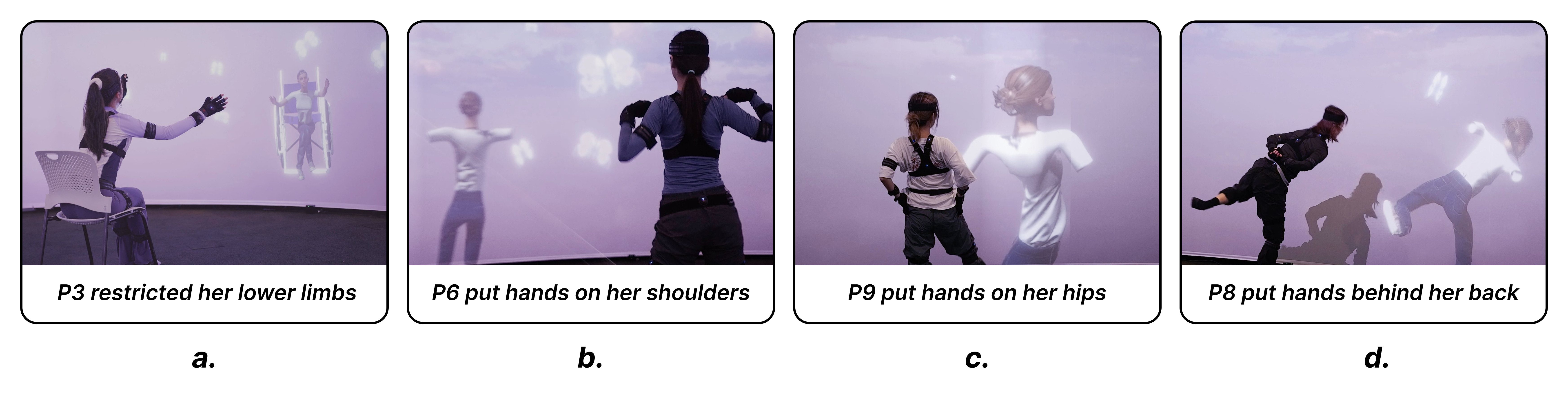}
  \vspace{-0.5cm}
  \caption{Self-restricted movement sets: \textbf{a.} Limiting the use of lower limbs when dancing with the avatar in a wheelchair; \textbf{b - d.} Limiting the use of upper limbs using different ways when dancing with the avatar without arms.}
 \label{restricted}
  \vspace{-0.3cm}
\end{figure*}

Dancers subconsciously restricted their own body parts due to avatar movement disability (the avatar in a wheelchair or the avatar without arms). While most dancers reacted to disabled avatars potentially with stereotypes, some also recognized and indicated the importance of challenging stereotypes to expand their movement possibilities. 

Dancers (P2, P3, P9, P11, P12, P14, P15) reported a strong sense of restraint given by the avatar in a wheelchair (see Fig. \ref{cv_limb}). For example, P2 felt the body of this avatar was completely different from her own body, restricting her freedom of movement. She \textit{"took the initiative to follow her rules and gave up the original body,"} limiting the use of lower limbs. She focused more on emotional expression instead of movements when performing with this avatar. Moreover, she rationalized the insensitivity of this avatar by saying, \textit{"She has been sitting in a wheelchair for a long time, and other parts of her body are not very flexible, which is actually normal."} P3 went through a strong emotional change while facing this avatar. She empathized with the avatar, imagining herself losing the ability to use her legs (see Fig. \ref{restricted} a). Some dancers like P9 and P15 even completely fixed their lower bodies during the improvisation. Just like them, many dancers self-restricted their lower body to different extents due to the image of the disabled avatar.

\begin{quote}
    \small\itshape
   "I was so sad that I almost cried... If you can't move, it is a big bad news for a dancer. In the beginning, my movements were more negative... I don't accept it, but later, I let it go and accept it, and I embrace it, just like embracing myself like embracing a butterfly, and I was thinking what kind of beautiful display I can still make." (P3)
    \end{quote}
    
Although dancers reported feeling less physically constrained with the avatar without arms, many still limited their upper limb movements in various ways. Some dancers placed their hands on their shoulders (P6), hips (P9, P12, P14, P15), or behind their backs (P1, P8, P13) to restrict their upper limb movements (see Fig. \ref{restricted} b - d). P4 mentioned that her arms entered a \textit{“follow state”} rather than a \textit{“lead state”}, even though they continued to move. Similarly, P10 noted reduced sensitivity in her upper limbs, which led her to move her arms less. These responses highlighted how the absence of visual representation for limbs could influence dancers’ movement decisions, even when there was no physical restriction.

However, some dancers (P9, P12, P14) recognized that their own preconceived notions about people with disabilities may have limited their movement choices. P9 remarked, \textit{"I think that my preconceived notions have indeed affected the design of some of my movements, including the legs... I feel that although her legs cannot move, she can actually move them, or fold them, or turn them more, but subconsciously, I just fixed my lower body, like I didn't have my lower body at all."} Similarly, P12 reflected that disabled individuals she encountered were often much stronger than societal expectations suggested. She noted that while some may lack physical limbs, they often possess greater physical strength, resilience, and mental or spiritual fortitude than those without disabilities. These reflections emphasized the importance of challenging stereotypes to allow for a more expansive approach to movement, even in the context of perceived physical limitations.

\subsubsection{Perceived freedom through avatar movement constraints} \label{freedom}

While dancers self-restricted their movements due to the disabled body parts of avatars, they (P1, P2, P6, P8, P9, P12, P13, P14, P15) specifically perceived freedom from the avatar without arms. Dancers reported deeper thoughts provoked by this avatar and movement explorations to push their physical limits and creative boundaries.

Dancers reported the experience of moving with the avatar without arms as intriguing and thought-provoking. This avatar, according to P2, was with \textit{"strong characteristics"} and impressed her. Although this avatar did not have upper limbs, she felt it \textit{"more liberating"} as she could focus even more on other body parts (e.g., legs and torso) and move them quite freely. Similarly, P6 found it fascinating to observe her movements through a disabled avatar, as they differed from her expectations, unlike the synchronized movements of healthy avatars. This mismatch sparked new choreographic ideas, as she noted, \textit{“When you lack something, you can find something else to make up for it… this imperfect thing can give you a different perspective.”} P13, while exploring more with her torso, also felt the need to exert greater effort, which moved her and made her feel full of vitality. Overall, the dancers’ experiences suggest that the limitations of disabled avatars can provoke deeper thoughts and challenge conventional approaches to choreography.

\begin{figure*}[htbp]
  \centering
    \includegraphics[width=.99\linewidth]{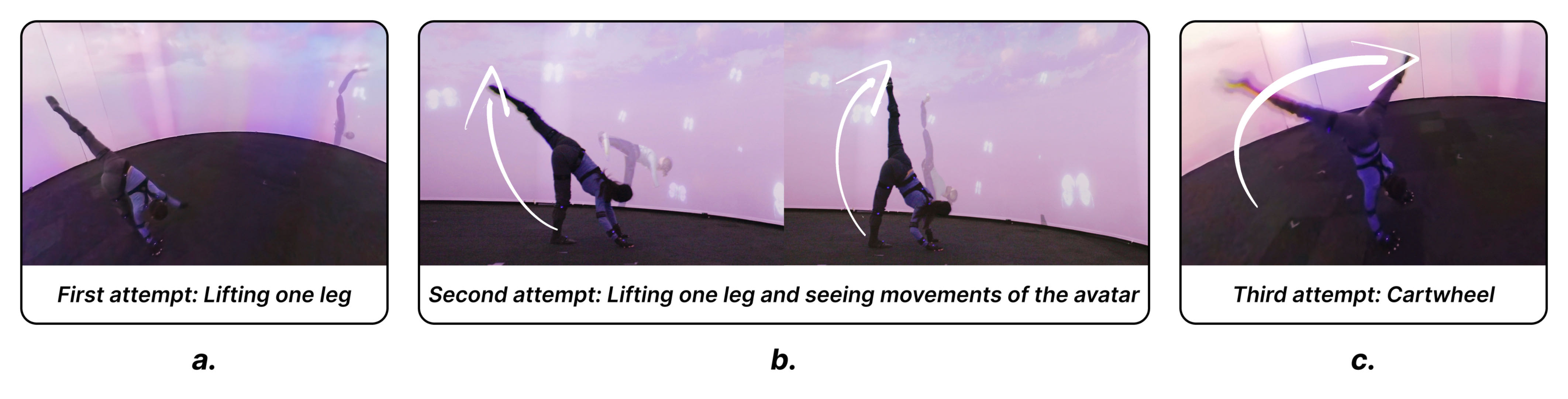}
  \vspace{-0.5cm}
  \caption{P6 experimented with the "cartwheel" movement three times to observe the avatar without arms.}
 \label{cartwheel}
  \vspace{-0.3cm}
\end{figure*}

\begin{quote}
    \small\itshape
   "The left shoulder seems to be a little insensitive. It may be a characteristic of the disability. When she twists her body, it is very... I felt that the alienation of the disabled person's body itself is very interesting, or very characteristic, or something worth showing off." (P2) 
    \end{quote}
As a result, dancers experimented with pushing their own physical limits and gaining new insights into choreography. P1 and P13, while self-restricting their upper limbs, explored the dynamic of balance and imbalance through weight shifts, tilting, falling, and rising. P6 attempted a \textit{"cartwheel"} but observed a \textit{"somersault"} from the avatar (see Fig. \ref{cartwheel}). This allowed her to achieve more difficult movements that she could not realize in real life, and with less effort. P9 pushed her body to the limits of its range, focusing on how the disabled upper limbs could guide her movements. She also experimented with shoulder-initiated movements, likening the sensation to a butterfly’s flutter. These explorations reveal how disabled avatars can inspire dancers to rethink movement possibilities and expand their creative boundaries.

\subsubsection{Shifting roles of avatars in improvisation} \label{shifting roles}
During the improvisation, participants had varying perceptions of the role of avatars. Some perceived avatars as extensions of themselves, allowing them to observe their own movements like audiences; some viewed avatars as their dance partners that were difficult to form connections and lacking in natural feedback. Some dancers also noted an interesting shift of dominance during the improvisation, with unfamiliar and disabled avatars often taking the lead. 

Dancers (P1, P3, P7, P8, P9, P11, P15) perceived avatars as extensions of themselves, enhancing their immersion and allowing them to become their own audience. P3, for example, felt that controlling the avatar in the virtual world enabled her to rehearse more efficiently, as she could direct the performance while observing herself from a distance. Similarly, P7 saw the avatar as a virtual self, playing different characters. P8 found the avatar provided a unique perspective by showing her own back, which differed from the usual front-facing view she got from a mirror. This allowed her to evaluate her movement quality in a new way. P9 likened the avatar to a shadow or a correlated version of herself on the screen, reinforcing the connection between her physical movements and the avatar’s performance. These experiences suggest that avatars offer dancers a fresh perspective on their movements, blurring the lines between self-perception and external observation.

Some participants (P2, P4, P6, P13) viewed avatars as dance partners, though forming a connection was challenging due to the lack of physical sensation. P6 saw the avatar as a separate entity that was always controlled by her, and felt responsible for considering its role in improvisation, much like with a human partner. P4, however, struggled with the avatar’s identity, questioning whether it was an independent being or just a projection. She expressed difficulty in acknowledging its existence, explaining that unlike a human partner, whom she could sense physically, she needed to see the avatar to recognize it. P4 reflected, \textit{“I completely regard the avatar as a new species... If it is a real person, I can really feel it with my body... I have to see the avatar before I know it exists, otherwise I will still be alone.”} Despite this confusion, she felt a need to care for the avatar, although the connection was vastly different from that with a human. As a result, she ultimately described the avatar as an \textit{“unharmonious partner.”} P13 similarly described the avatar as an untouchable partner in the digital space, perceptible only through sight. These reflections might suggest that the lack of sensory feedback made it hard for dancers to form meaningful connections with avatars just as human beings.

Furthermore, avatars often failed to provide dancers with the real-time, responsive interactions that human partners offer. P4 observed that the reactions from avatars were sometimes \textit{"programmed and objective with lags,"} requiring her to carefully monitor their movements, as avatars could not convey feedback independently. In contrast, human partners respond naturally through sound or physical sensations, even when out of sight. P2 similarly felt a disconnection, explaining that she found it difficult to \textit{"know its thoughts"} as she would with a human partner. Occasionally, avatars would \textit{"seem stuck,"} unable to respond to her \textit{"invitation to come here together."} P15 highlighted the limited unpredictability of avatars compared to human partners, noting that this reduced randomness restricted creative exploration. The lack of spontaneity and natural feedback from avatars presented a challenge to creative improvisation, though it also prompted dancers to adapt in novel ways.

Interestingly, participants observed the switch of dominance during the improvisation with avatars. P6 noted that her sense of control depended on her visual connection to the avatar: when she couldn’t see the avatar, she felt in control, but when she turned around, she was drawn to the avatar’s movements, allowing it to lead her improvisation. She preferred the latter situation when the avatar took the lead, giving her unfamiliar things to imitate. For P2, different avatars gave her a sense of control to a different extent. For example, avatars that were disabled and much more different from her own body (avatars in a wheelchair and without arms) took the dominance and controlled her movements, making her \textit{"try to find a way to be more like the appearance of the avatar."} On the contrary, she reached "a relatively equal state of cooperation" with healthy avatars that more closely resembled her own body. Consistent with P2, P12 and P14 also felt led by disabled avatars during improvisation, while they assumed control when interacting with other avatars. These experiences suggest that the level of physical difference between the dancer and the avatar influences the power dynamic, with unfamiliar or disabled avatars often taking the lead.

\section{Discussion}\label{sec:Discussion}
Our study investigated how avatars of different genders, shapes, and physical limitations affect dancers' movements (RQ1) during improvisation and how dancers perceive their interactions with avatars in immersive situations (RQ2). 
Through an embodied approach, avatars challenged dancers to see themselves and other identities in new ways, breaking free from habitual movement patterns. The results reveal that engaging with gendered or disabled avatars not only encouraged physical experimentation but also fostered critical self-reflection. Dancers reported feeling free to explore gendered or constrained motions, confronting biases and gaining insight into diverse identities, although focusing on avatars sometimes disrupted continuity in their performance. By embodying physical limitations or opposing gender traits through the avatars, participants experienced a shift in perspective, enabling them to empathize with others’ experiences as if viewed through their own bodies. This highlights avatars’ potential as tools for exploring physical identity and fostering inclusivity in movement practices.

\subsection{Supporting Movement-Based Representation of  Underrepresented Groups in Dance} \label{representation}

To our knowledge, this is the first study to engage dancers in representing their bodies with avatars of underrepresented body types in dance. With avatars, they were able to be empathetic from an embodied experience instead of just pretending to be another person with disability or non-normative body shapes, fostering an empathy-building process of underrepresented groups from a dance perspective. Through embodied practice with avatars of different genders, body shapes, and physical limitations, dancers reported shifts in perspective on underrepresented identities, from moving with stereotypes to breaking free from negative stereotypes towards different identities.

Despite the growing awareness of the importance of including representations of various identities both in the physical and digital world, it is still hard for healthy and normative people to be truly empathetic with people of different demographics, leading to negative stereotypes that could result in underperformance both physically and mentally \cite{yee2006walk}. For example, researchers have found gender stereotypes in dance, where dance involving softness in movements could be considered as a female motor activity, undermine dance performance and motor learning in boys \cite{bastos2023gender}. While digital spaces provide a unique opportunity to bring different identities on stage through customizable and easily updated avatars, people with disabilities still often encounter negative stereotypes as in real life, such as disabled people can only sit in a wheelchair and cannot move \cite{mack2023towards}. Our findings also revealed dancers' stereotypes towards their opposite gender and disabled people at the beginning, limiting their creative possibilities. Therefore, although researchers set out to explore how different identities want to represent themselves in digital spaces \cite{zhang2022s}, negative stereotypes can still exist to hinder the genuine inclusion of various groups due to the lack of empathy in normative people. However, dancers who dance should be of all genders, physical identities, and body shapes.

Our study, instead of probing the way underrepresented groups in dance represent themselves, provides normative dancers with a chance to move into a different identity's body, allowing them to take the perspective of another person directly for deeper empathy and understanding. Through an embodied movement-based approach, our dancers expressed their thoughts through movements and, from movements, learned the limits and stereotypes. Therefore, a gradual shift of perspective occurred in dancers during the improvisation. From seeing avatars of different identities and conforming to stereotypes to moving out of the scope of them, dancers were aware of potential stereotypes in their deep minds. For example, they reflected that people sitting in a wheelchair could be stronger than healthy people and have various ways of moving. Avatars functioned as catalysts, gradually increasing the visibility of underrepresented bodies in dance and leading to respectful movement exploration against established impressions. This further suggests the potential of using avatars of underrepresented body types as sources of inspiration for movement artists.

\subsection{Creativity Support through Embodied Interaction } \label{creativity support}

\subsubsection{Differed defamiliarization effects of avatars as extended self or partners} 

Our findings revealed that avatars could augment natural improvisation as defamiliarization tools, changing the way dancers perceive their own bodies and thus influencing their movements (Section \ref{defamiliarization}). Avatars, either perceived as the extension of self or dance partners, distanced dancers from their movement habits and opened up new ways of moving (Section \ref{shifting roles}). This result aligns with previous research \cite{zhou_here_2023} on MR mirror and humanoid avatar-supported improvisational dance making. However, while researchers suggest that avatars allow dancers to defamiliarize with themselves by observing themselves objectively like someone else \cite{zhou_here_2023}, our dancers perceived avatars as the extension of themselves in the virtual world or their dance partners.

For dancers who perceived avatars as an extension of themselves in digital space, the avatars served as a medium for stepping into different characters, triggering an internal psychological shift that reinforced natural improvisation. This aligns with Pina Bausch’s approach, which emphasizes not how people move but what moves them \cite{climenhaga2008pina}. In her view, dance is an innate human expression that emerges from life experiences rather than mere technique \cite{climenhaga2008pina,zhou2021dance}. Our findings suggest that avatars, as a technological tool, have the potential to evoke dancers’ inner thoughts, guiding movement in an organic and intuitive way. While prior work has highlighted the tension between dance as an expressive art form and as a quantifiable technique \cite{zhou2021dance}, our study suggests that avatars may help bridge this divide by shaping dancers’ mindsets and supporting the meaning-making process in improvisation.

For other dancers who viewed avatars as their dance partners, however, the defamiliarization effect was found to be restrained due to the lack of physical sensations and effective feedback. Improvisation with a partner, which is often called contact improvisation, can facilitate strong defamiliarization based on constant unknown sensory feedback from the partner \cite{carlson2019shifting}. On the contrary, avatars could only be sensed by dancers' eyes and only give feedback when being observed. The lack of physical sensations and constant feedback made dancers hard to establish connections with avatars, reducing the level of defamiliarization that could be provided by human partners.

\subsubsection{Disrupting stereotypes and norms through constraints}

Our findings show that dancers took the physical limitations and technological constraints brought by avatars as a chance to open up kinaesthetic creativity (Section \ref{creativityfromconstraints} and Section \ref{freedom}). For example, P6 experimented with the "cartwheel" to observe a "somersault" on the avatar without arms; P9 reported gaining inspiration through the avatar distortion. While dancers might still exhibit certain stereotypes when interacting with avatars, some started breaking the negative stereotypes that limited their bodily interactions.

Many previous works set deliberate boundaries to push dancers out of usual movement patterns \cite{warburton2018imagine,alaoui2021rco}. For example, Radical Choreographic Object is a participatory performance that introduces social and technological constraints into choreography to disrupt social norms by forcing participation \cite{alaoui2021rco}. Participants were found to have an equivocal reaction to the disruption, firstly conforming to social norms and predetermined everyday rituals, and then progressively behaving outside of acceptable frames. Moreover, the technological constraints involved exhibit two polarities, both oppressive and expressive, delimiting dancers' behaviors while providing a space for less constrained expression in the meantime. 

In line with previous research \cite{alaoui2021rco}, our dancers exhibit similar patterns while interacting with avatars with physical disabilities and technological constraints. In the beginning, dancers overly restricted themselves due to negative stereotypes of people with disabilities, which largely undermined their creative exploration \cite{yee2006walk}. Later on, they started breaking free from these limits, exploring different ways of moving in a disabled body and testing their own limits. The technological constraints, such as avatar distortion and failure to show details, inspired dancers to change their own movements in unexpected ways by restricting their expression to a certain scope. This suggests the possibility of introducing constraints by incorporating various avatars into dance practice, helping dancers break free from daily rituals and create from a new perspective.

\subsubsection{Less constrained environment for identity exploration}

Our findings show that dancers saw the improvisation with avatars of different genders in the immersive environment as a chance to explore their gender identity, leading to experiments with gender-stereotyped movements (Section \ref{gender identity}).  For example, P12 did male postures in Chinese classical dance, while P14 did feminine movements, including a lot of body curves. Interestingly, P6 experimented with feminine movements with the male avatar to observe a funny effect. As P4 mentioned, gender identity seems to hold significance in dance while she continually questioned the reason. Dancing with avatars, however, gave her a less constrained environment to test different gendered movements and explore her own gender identity. 

This finding is in line with prior findings on avatars and gender identity \cite{freeman2022acting,zhang2024gender}. It has been found that some non-binary dancers would avoid discrimination and violence by hiding their gender identity and disguising themselves to be cisgender people \cite{sonik2023embodied}, keeping their identity "invisible." Avatars in social VR, however, were considered to enhance the embodied visibility of non-cisgender individuals by facilitating the exploration, experiment, and expression of their identity online \cite{freeman2022acting}. Our findings suggest that avatars may benefit not only non-cisgender individuals seeking places for less constrained identity expression, but also those who are confused and exploring possibilities about their identities, such as P4, providing a relatively free space for identity exploration. In this space, people could explore diverse aspects of themselves without facing judgment and discrimination that frequently occur in the real world \cite{zhang2024gender}, rather than submissively accept cisgender stereotypes within societal norms \cite {nowak2005influence}.

\subsection{Drawbacks Limiting Creative Possibilities}

\subsubsection{Loss of agency due to technological constraints} \label{loss of agency}

The results show that dancers could be distracted when paying too much attention to avatars, resulting in failure to perceive their own bodies and less vivid and three-dimensional movements (Section \ref{distraction}). Moreover, the wheelchair condition was reported as less resembling dancers' own bodies (Section \ref{sec:embodiment}) and largely restricted their movement exploration scope (Section \ref{self restrict}). 

We relate this finding to prior work on kinaesthetic creativity in dance, where the dance creativity was connected with the ability to shift between various interactive visuals \cite{hsueh_understanding_2019}. It was suggested that absorbing technological tools and transforming them into dancers' bodies could be viewed as a type of skill acquisition \cite{svanaes2013interaction}. However, users might lose agency if the process is not handled well, resulting in more passive instead of self-initiated and exploratory movements \cite{hsueh_understanding_2019}. 

We hypothesize that the distraction given by avatars might derive from the difficulty in communicating with them, leading to unfavorable skill acquisition process \cite{svanaes2013interaction}. As our dancers noted, the avatar could only be sensed when in sight, unlike human partners, who could be sensed through the quality of touch, temperature, gaze, heartbeat, etc. This made the process of getting familiar and establishing connections with avatars harder. We suggest that future work should incorporate multi-sensory experience into the use of avatars for creativity support, alleviating the difficulty in transforming avatars into dancers' own practice.

We speculate that the restriction brought by the wheelchair condition also relates to the loss of agency. Although the statistics did not show the significance of the wheelchair condition on \textit{Agency}, the significant value on \textit{Total embodiment} suggests that dancers felt the avatar in a wheelchair less resemble themselves, making them have less control of the situation. Dancers also reported in interviews that their movements were significantly influenced and felt restricted to a large extent, which is consistent with previous research suggesting a greater similarity between the avatar and user enhanced embodiment effect \cite{latoschik2017effect}. In this case, the physical constraint designed for creativity support led to negative impacts instead \cite{hsueh_understanding_2019}. Therefore, we suggest using avatars with higher embodiment effects in co-creative dance systems, giving dancers agency to perform self-initiated movement exploration.

\subsubsection{Lack of expressiveness in realistic humanoid avatars}\label{expressiveness}

In contrast to earlier findings, which suggest avatars could convey not only movements but also character qualities \cite{raheb_choreomorphy_2018}, our dancers noted the lack of expressiveness in avatars, limiting the possibilities of creative expressions (Section \ref{limited expressiveness}). We speculate that this inconsistency is rooted in the different designs of avatars. In prior work, avatars were made to be abstract or cartoonistic \cite{raheb_choreomorphy_2018}, while we selected realistic avatar sets. Through prior work where dancers improvised with abstract visual representations of their body contours, researchers have suggested the abstract nature in kinaesthetic creativity \cite{hsueh_understanding_2019}. Dancers naturally embrace the abstract and form meaningful relationships with collaborative performing agents \cite{zhou2021dance}. Our findings suggest that the use of avatars also follows previous suggestions of using abstract instead of accurate mapping of MoCap to explore expressive movement qualities. Future work should either incorporate multi-sensory feedback to make realistic avatars more expressive like human beings, or choose more abstract avatars to investigate how they facilitate artistic expression.

\subsection{Design Implications}

Based on our findings, we propose guidelines for future dancers, performers, researchers, and developers:

\begin{enumerate}
    \item For dancers, we advocate incorporating avatars of various identities and bodies into their daily practice for embodied creativity support and to counter negative stereotypes that can limit the performance and creativity \cite{yee2006walk}, including underrepresented bodies in dance through an embodied movement-based approach. As highlighted in Section \ref{defamiliarization}, avatars led dancers to distance themselves from habitual movements through the defamiliarization effect. Furthermore, Section \ref{creativityfromconstraints} and \ref{freedom} reveal that the physical and technical constraints of avatars even brought dancers with more creative possibilities, leading them to discover and combat stereotypes towards different identities in real life. Additionally, our findings in Section \ref{gender identity} reveal how dancers took the improvisation with avatars as a less constrained chance to explore their own identity, uncovering more ways of moving through an embodied approach.
    \item For performers using MoCap in live performances, they may face calibration challenges that would interrupt their performance, as we found in Section \ref{creativityfromconstraints}. Before the performance, they should receive training to familiarize themselves with the specific constraints and opportunities of performing with MoCap to minimize errors. During the performance, we suggest doing pre-calibration. Dancers could also incorporate calibration procedures into their choreography, optimizing artistic expression while maintaining performance integrity when recalibration is required—an approach already explored in recent MoCap live performances \cite{noauthor_zelia_nodate}. 
    \item To address the distraction brought by avatars that are unfamiliar and hard to communicate with (Section \ref{distraction}), we propose incorporating multi-sensory feedback in realistic avatars. As dancers noted, they were distracted by avatars and could not receive feedback; they would also be distracted by human partners in their daily practice, but this could be mediated by feedback like physical sensation (Section \ref{shifting roles}). Therefore, feedback could partially alleviate the distraction brought by avatars. By incorporating multi-sensory feedback, realistic avatars would be closer to human partners, allowing dancers to allocate their attention properly.
    \item To address the loss of agency related to the wheelchair condition (Section \ref{sec:embodiment} and \ref{self restrict}), we suggest that  future research should allow dancers to move the wheelchair to give them more agency. As detailed in Section \ref{self restrict}, dancers felt a strong sense of restraint and almost fixed their lower limbs, which greatly restricted their creative expressions. By allowing them to move in a wheelchair, they would gain more freedom in movement possibilities and more agency over their own body.
    \item  To address the limited expressiveness of avatars (Section \ref{limited expressiveness}), we propose incorporating facial expressions in realistic avatars or changing rhythms when using avatars. In Section \ref{limited expressiveness}, dancers reported that avatars could hardly show emotional depth out of only movements. Incorporating facial expressions in avatars, although it could not solve all the expressive issues like movement qualities, would add much more emotional feedback. Nevertheless, adding facial expressions could potentially lead to more distraction. In line with prior work exploring performer expressivity in distracting environments \cite{LC2023contradiction}, the avatar design should carefully balance expressiveness and distraction. Therefore, changing music when performing with avatars would also be a way to increase expressivity while maintaining minimal distraction, as dancers may have been used to dancing with music.
    \item Another solution to address the limited expressiveness would be using more abstract avatars (Section \ref{limited expressiveness}). As kinaesthetic creativity has an abstract nature \cite{hsueh_understanding_2019}, dancers found it natural to embrace the abstract instead of accurate mapping of MoCap to explore expressive movement qualities \cite{zhou2021dance}. Using more abstract avatars would be meaningful for dancers to form relationships with performing agents and improve expressivity.
    \item For developers, we suggest developing MoCap systems that are more stable in live environments, avoiding constant recalibration, which would largely distract performers. Besides, calibration systems that can faster recalibrate and are easier to operate during the show are needed for live performances.

\end{enumerate}

\subsection{Limitation}

\subsubsection{Design of avatars} 

We selectively included only avatars of heavyset, gender opposition, and certain physical disabilities, excluding those related to other body sizes, non-binary genders, or invisible disabilities \cite{mack2023towards}, which could limit the exploration of a broader range of embodied experiences. Although heavyset body shapes were incorporated, the changes were subtle and less noticeable, which might result in inconspicuous embodiment effects and improvisation influences. Additionally, the available avatar assets \cite{noauthor_character_nodate} contained gender stereotypes, particularly in the body shapes associated with different genders, which could influence dancers' perception and movement choices outside of the gender change itself. Moreover, the avatar design was predominantly “Asian,” potentially impacting the generalizability of the findings across diverse ethnic groups. Future studies could address these issues by including avatars of different ethnic groups identified by ethnic clothes, additional avatar traits like limb length, body sizes, non-binary genders, and invisible or fluctuating disabilities into the embodied interactions, to see whether and how these features can influence movement exploration. 

The fixed viewpoint of the avatar in the virtual environment posed another limitation. Participants viewed the avatar from a single perspective, either mirrored or front-to-back, which is not typically how dancers interact with their human partners. Providing more customizable and dynamic viewpoints, such as side-to-side, top-down, or even a 360-degree perspective, would allow dancers to engage with avatars in a way that more closely mirrors real-world interactions, enriching movement experiences in both virtual and hybrid dance practices.

\subsubsection{Demographics}


\paragraph{Limited gender diversity}
The generalizability of our results is constrained by participant demographics, as our sample included limited male dancers and no non-binary dancers. A more diverse group of dancers might have provided richer insights into how gender identity influences the use of avatars in dance. For instance, we observed that participants did not engage more in jumping or turning movements with the avatar without arms. However, this finding might differ with more male participants, who might perform such movements more frequently due to the muscular physiques often characteristic of male dancers \cite{clegg2016cool}. It will then suggest that male dancers’ interactions with the no-arm avatar might diverge from those of female dancers in this specific context. 

Moreover, non-binary identities have long been underrepresented onstage in dance, leading to non-binary dancers often needing to conform to strict male and female representations, or ignore gender identities like in modern dance \cite{wiley2020designing,new_adventures_matthew_2024}. Some transgender people even hide their identities to avoid discrimination and violence \cite{sonik2023embodied}.
However, if we have non-binary dancers to improvise with avatars of different genders in a relatively private space, they may react in a different way compared to their daily practice and their binary fellows. For example, while binary dancers might align their movements with an avatar’s perceived “masculine” or “feminine” traits, non-binary dancers could blend or subvert these characteristics, creating hybrid movement vocabularies. This perspective might lead to richer interpretations of how gendered and non-gendered avatars could influence individualized expressions, bringing often marginalized  non-binary dancers onstage. 

\paragraph{Lack of physical diversity} 
While our study was designed for non-disabled dancers to see other people’s limitations from their own bodies through an embodied approach, including people with disabilities would offer valuable insights. 
Prior research has shown that embodied performance with avatars facilitates the discovery of alternative selves and fosters emotional well-being by enabling the expression and sharing of personal experiences \cite{ryu2023embodied}. Similarly, dancers with disabilities may find embodied improvisation with avatars a powerful means of exploring alternative self-representations, leading to emotional release and creative growth.
Future work could expand this research to include dancers with disabilities in co-designing avatars that authentically represent their needs and preferences. Additionally, studying how disabled dancers use avatars to explore new movement possibilities—by overcoming physical limitations or challenging social perceptions of disability—would offer valuable insights into how digital spaces reshape notions of embodiment, identity, and artistry in dance.

\paragraph{Homogeneity in dance experience and age range limitation}
While our study focused on professional dancers performing MoCap-supported live performances—tasks typically suited to individuals with extensive training and within a younger age range—we did include P13 (age 47), P15 (age 34), and P12 (age 32) in our study. However, we noticed that including novice dancers in this study may uncover additional insights into dance learning. Prior research has explored how avatars can enhance dance learning \cite{cisneros2020wholodance}, showing that using dissimilar avatars can improve novice dancers’ skills by encouraging diverse movement exploration \cite{fitton2023dancing}. Including novice dancers in future studies could provide valuable insights into how avatars with different affordances influence learning processes, creativity, and engagement in dance. This approach could also broaden the applicability of our findings to education and amateur dance practices.

\paragraph{Cultural and ethnic diversity}
Similar to previous work \cite{wallace2023embodying,fairlie2023encouraging,trajkova2024exploring}, we also used a specific population. Our participants are predominantly from Chinese-speaking backgrounds, although they have different dance genres like ballet, modern dance, Chinese dance, swing, etc. 
However, dance, as a form of bodily “text,” is deeply embedded in cultural and historical contexts, giving rise to distinct genres with unique movement languages \cite{desmond1993embodying}. Including dancers from varied cultural and ethnic backgrounds would significantly enrich our data. For instance, certain dance styles, such as Indian Dance featured in a previous MoCap live performance \cite{noauthor_luyang_nodate}, have unique cultural, symbolic, and movement-based elements that may affect how dancers engage with avatars. 
Involving Western dancers in our study may specifically influence our findings regarding gender identity exploration. As suggested by prior work \cite{gutierrez2020gendered}, different systems of thought of East Asian and Western cultures might influence their gender identity construction and expression. Masculinity and femininity can co-exist in an East Asian individual, while a Western individual can only be either masculine or feminine. Therefore, Western dancers may strictly adhere to their own gender while performing as the opposite gender, as opposed to our participants who did gender-stereotyped movements based on the avatar's gender. Another possibility would be that Western dancers might perform just as the avatar's gender from the beginning to the end, while our participants, like P6, experimented with feminine movements with the male avatar to see the funny effect. Including Western dancers in future research would shed light on whether responses to avatars differ between cultures, enriching the research of identity and representation in dance.

\subsubsection{Calibration problems in live performances}

Although calibration was performed prior to the live performance in the second stage, occasional recalibration was necessary during the performance, disrupting its continuity. Intense movements sometimes caused straps or sensors to slip, leading to avatar distortion. Additionally, severe distortions occasionally occurred due to synchronization issues with the software. These technical challenges highlight certain limitations of using MoCap technology in live performances. Future research should focus on engaging performers in real MoCap-supported live scenarios to comprehensively identify and address these issues. Exploring solutions such as improved sensor stability, real-time error correction, and advanced synchronization algorithms could significantly enhance the reliability of MoCap systems in dynamic performance settings.

 \subsubsection{Study design}

The within-subject design may have led to a practice effect, where participants became more familiar with the avatars over time. Although we have counterbalanced using Latin Square Design, this could still result in unequal engagement, as dancers may have explored later avatars more thoroughly due to increased comfort with the system. Consequently, participants might have developed a stronger connection to certain avatars, not based on their design, but due to the order of exposure. Randomizing avatar interactions in future studies could help reduce this bias.

\subsubsection{Self-reported survey data}

One of the limitations of this study was our survey relied on self-reported data, which is inherently subjective and constrained by the specific questions we asked. For instance, participants may underreport or overestimate the embodiment effect of avatars, leading to inconsistent or incomplete data. Future studies can incorporate more objective metrics (e.g., sophisticated motion-tracking measures) to yield more reliable and detailed insights into how dancers experience avatars.

\subsection{Future Work}

This study demonstrated how avatars with diverse physical conditions influence dancers’ movement (RQ1) and perception (RQ2), providing insights into embodied experiences that prompt them to reconsider habitual movement patterns and physical identities. However, several areas warrant further exploration.

First, our findings suggest both the facilitating and distractive effects of avatars on improvisation. While avatars encouraged defamiliarization and experimentation, they also introduced challenges such as distraction and loss of agency, depending on the context, avatar scenario, and the dancer’s focus. Future research could refine avatar designs to address these challenges by integrating multi-sensory feedback or facial expressions to enhance realism, allowing dancers to connect with avatars more naturally—similar to human partners. Alternatively, abstract avatars could be explored to align with the abstract nature of kinaesthetic creativity while minimizing cognitive overload.

Second, expanding participant diversity would further enrich the study’s implications. Future studies could include dancers with disabilities and those from varied cultural or training backgrounds to explore how embodied virtual experiences translate across different populations. Longitudinal research could also examine how prolonged engagement with avatars shapes dancers’ self-perception and movement styles over time.

Third, alternative experimental designs could extend our findings. One approach could involve interactive duet scenarios where dancers co-move with AI-driven avatars, examining how real-time responsiveness influences improvisation. Another possibility is integrating biometric data (e.g., heart rate, muscle activation) to assess physiological responses to avatar-induced movement shifts. Additionally, research on audience perception of avatar-mediated performances could offer valuable insights into how virtual embodiments impact spectators’ interpretations of dance.

By addressing these areas, future research can further establish the role of avatars as tools for creativity, identity exploration, and inclusivity in movement-based practices.

\section{Conclusion}\label{sec:Conclusion}
Through semi-structured interviews, self-reported surveys, and computer vision data collected from 15 non-disabled professional dancers, we investigated how avatars with underrepresented bodies in dance—such as heavyset figures, gender-opposed representations, and those with disabilities—impact movement and perception in practice and performance settings. Our findings offer insights into how dancers interact with avatars that challenge their habitual movement patterns and physical identities. We then propose design solutions to mediate the facilitating and disruptive effects of avatars and the challenges of using MoCap in live performance settings. These findings emphasize the potential of avatars to act as tools for reimagining physical identity and advancing inclusivity in artistic practices, highlighting their value not only in redefining dance improvisation but also in fostering deeper connections across diverse bodies and identities.

\begin{acks}
We sincerely thank all dancers who contributed to this research, including Shi Yingnan, Cai Yuchen, Yang Yi, Katrina Zeng Yulin, Lo ShuMan, Wang Tianhao, Wen Yingxi, Xiaowen, Lydia Li Yanheng, Leung ChungMan, Wen Feiyang, Malu, Kylie Winona Choy, Yu Guoming, and Chau MengHan. Special thanks to Louise for assisting with avatar and virtual scene development and to Sky Suen and Leoson Cheong for their support in mapping the virtual environment to the 360-degree space. This work was supported by the Research Grants Council Theme-based Research Scheme [T45-205/21-N] and the Hong Kong Research Grants Council General Research Fund [9043630].

\end{acks}

\bibliographystyle{ACM-Reference-Format}


\newpage
\appendix

\label{sec:Appendix}
\section{Interview Questions \label{appendix}}

\label{interview}

\begin{itemize}
    \item \textbf{Prior experience with avatars}
\end{itemize}

Do you have prior experience with avatars?

\begin{itemize}
    \item \textbf{Improvisation with avatars}
\end{itemize}

Can you describe your experience of improvisation with different avatars (ask based on their feedback in the survey)?

What movements could you make or intend to make due to each avatar?

Did you notice anything that influenced your decision-making process during improvisation? How?

If so, can you share any instances where the visual feedback influenced your improvised movements?

\begin{itemize}
    \item \textbf{Perception of interactions with different avatars}
\end{itemize}

How did you feel when you first saw your movement visualized through an avatar in real time?

Did the appearance of the avatar (different affordances) affect how you perceive your movements? If so, how?

Did avatars influence your feelings or emotions during dance improvisation?

How did the use of an avatar affect your sense of ownership over your body? 

Did you feel more or less in control of your movements?

What were the roles of avatars to you? 

\newpage
\section{Results of Individual Survey Questions} 
\label{survey question}

\begin{table}[h!]
\centering
\caption {\textit{P-value} Comparison of Normative (N) and Conditions (C1 - C4) on Individual Survey Questions. \label{question} }
\small
\renewcommand{\arraystretch}{1.2}  
\begin{tabular} {p{4cm}cccc}
\toprule
\textbf{Question}  &\textbf{C1} &\textbf{C2}  &\textbf{C3}   &\textbf{C4}    \\
\midrule
Q1. I felt as if the virtual body was my body. 
   & 0.164     & \textbf{0.029}    & \textbf{0.021} & 0.123   \\
Q2. It felt as if the virtual body I saw was someone else.   & 1.000     & 0.103    & 0.084 & 0.357   \\
Q3. It seemed as if I might have more than one body.
 & 0.454     & 0.432    & 0.328 & 0.370   \\
Q4. I felt as if the virtual body I saw when looking at the screen was my own body.  & 0.313     & 0.683    & \textbf{0.008} & 0.096   \\
Q5. I felt as if the virtual body I saw when looking at myself on the screen was another person.   & 0.582     & 0.398    & 0.787 & 0.571   \\
Q6. It felt like I could control the virtual body as if it was my own body.
 & 0.075     & 0.334    & \textbf{0.001} & 0.058   \\
Q7. The movements of the virtual body were caused by my movements.
   & 0.384     & 0.334    & 0.192 & 0.164   \\
Q8. I felt as if the movements of the virtual body were influencing my own movements.   & 0.885     & 0.349    & 0.525 & \textbf{0.037}   \\
Q9. I felt as if the virtual body was moving by itself.  & 0.565     & 0.774    & 0.779 & 0.415   \\
Q10. I felt as if my body was located where I saw the virtual body.
 & 0.223     & 0.890    & 0.259 & 0.215   \\
Q11. I felt out of my body.  & 0.199     & 0.619    & 0.597 & 0.723   \\
Q12. I felt as if my (real) body were drifting toward the virtual body or as if the virtual body were drifting toward my (real) body.  & 0.510     & 0.308    & 0.510 & 0.737   \\
Q13. It felt as if my (real) body were turning into an ‘avatar’ body.
 & 0.779     & 0.469    & 0.709 & 0.510   \\
Q14. At some point it felt as if my real body was starting to take on the posture or shape of the virtual body that I saw.   & 0.164     & 0.265    & 0.444 & 0.301   \\
Q15. At some point it felt that the virtual body resembled my own (real) body, in terms of shape, skin tone or other visual features.   & 0.370     & 0.589    & 0.764 & 0.719   \\
Q16. I felt like I was wearing different clothes from when I came to the laboratory.   & 0.334     & 0.250    & 0.607 & 0.152   \\
\bottomrule
\end{tabular}
\Description{\textit{P-value} Comparison of Normative (N) and Conditions (C1 - C4) on Individual Survey Questions.}
\end{table}

\end{document}